\documentclass[sigconf,10pt,screen]{acmart}

\acmSubmissionID{123}
\renewcommand\footnotetextcopyrightpermission[1]{}

\usepackage[english]{babel}
\usepackage{blindtext}
\usepackage{outlines}
\usepackage{subfigure}
\usepackage{caption}
\usepackage{subcaption}
\usepackage{algorithm}
\usepackage{algorithmic}
\usepackage{amsmath}
\usepackage{titlesec}
\usepackage{cleveref}
\crefformat{section}{\S#2#1#3}
\usepackage{xurl}
\usepackage{enumitem}
\usepackage{soul}
\usepackage{wasysym}
\usepackage{caption}
\makeatletter
\renewcommand\subsubsection{\@startsection{subsubsection}{3}{\z@}%
  {-0.5\baselineskip \@plus -2\p@ \@minus -.2\p@}%
  {.25\baselineskip}%
  {\ACM@NRadjust\@subsubsecfont}}
\makeatother

\renewcommand\footnotetextcopyrightpermission[1]{} 
\setcopyright{none}

\newcommand{\rname}{{\sc Heron}\xspace}
\newcommand{\cloudname}{{Azure}\xspace}

\newcommand{\parab}[1]{\vspace{0.03in}\noindent{\bf #1}}

\newcommand{\fix}{\textcolor{red}}

\titlespacing\section{2pt}{1pt plus 1pt minus 1pt}{1pt plus 1pt minus 1pt}
\titlespacing\subsection{2pt}{1pt plus 1pt minus 1pt}{1pt plus 1pt minus 1pt}
\titlespacing\subsubsection{2pt}{1pt plus 1pt minus 1pt}{1pt plus 1pt minus 2pt}
\newcommand{\subheading}[1]{\vspace{2pt}\textit{#1}}

\definecolor{lightgray}{gray}{0.92}
\newcommand\greybox[1]{%
\vspace{6pt}%
      \par{\centering\colorbox{lightgray}{%
              \begin{minipage}{3.3in}#1\end{minipage}%
                        }%
                                    \vskip 2pt%
                                    \vspace{1pt}%
                                                }}

\acmDOI{}

\acmISBN{}

\acmPrice{}
\expandafter\newcommand\csname r@tocindent4\endcsname{4in}
\begin{document}
\title{AI Greenferencing: Routing AI Inferencing \\to Green Modular Data Centers with \rname}



\renewcommand{\shortauthors}{X.et al.}

\author{%
  \fontsize{10pt}{12pt}\selectfont{}Tella Rajashekhar Reddy, 
  \fontsize{10pt}{12pt}\selectfont{}Palak,
  \fontsize{10pt}{12pt}\selectfont{}Rohan Gandhi,
  \fontsize{10pt}{12pt}\selectfont{}Anjaly Parayil,
  \fontsize{10pt}{12pt}\selectfont{}Chaojie Zhang,
  \fontsize{10pt}{12pt}\selectfont{}Mike Shepperd,\\
  \fontsize{10pt}{12pt}\selectfont{}Liangcheng Yu,
  \fontsize{10pt}{12pt}\selectfont{}Jayashree Mohan,
  \fontsize{10pt}{12pt}\selectfont{}Srinivasan Iyengar,
  \fontsize{10pt}{12pt}\selectfont{}Shivkumar Kalyanaraman,
  \fontsize{10pt}{12pt}\selectfont{}Debopam Bhattacherjee
}
\affiliation{%
\vspace{0.05in}
  \fontsize{10pt}{12pt}\selectfont{}Microsoft
}

\begin{abstract}
    \noindent AI power demand is growing unprecedentedly thanks to the high power density of AI compute and the emerging inferencing workload. On the supply side, abundant wind power is waiting for grid access in interconnection queues.
    
    In this light, this paper argues bringing AI workload to modular compute clusters co-located in wind farms. 
    Our deployment right-sizing strategy makes it economically viable to deploy more than $6$~million high-end GPUs today that could consume cheap, green power at its source.
    We built \rname\footnote{Heron of Alexandria~\cite{heron} invented wind wheel that could power a machine.}, a cross-site software router, that could efficiently leverage the complementarity of power generation across wind farms by routing AI inferencing workload around power drops. Using $1$-week of \textit{coding} and \textit{conversation} production traces from \cloudname and (real) variable wind power traces, we show how \rname improves aggregate goodput of AI compute by up to $80\%$ compared to the state-of-the-art.
\end{abstract}

\maketitle
\pagestyle{plain}

\section{Introduction}
\label{sec:intro}

\noindent With the rise of transformers~\cite{attention_is_all_you_need} and large language models (LLMs)~\cite{gpt4, gpt3_5, llama3, llama2, gemini}, and the rapid adoption of AI~\cite{peng2023study, ms365_copilot, github_copilot, bick2024rapid} by the industry, governments, and individuals, it is now paramount to satiate the growing energy needs of AI. 
Reports~\cite{goldman_sachs_may24, luccioni2024power} indicate a GPT query consumes far more energy than a web search, and AI image generation rivals smartphone charging. 
GPUs running AI workloads 
are getting significantly more power dense~\cite{power_density_1, power_density_2, power_density_3, nvidia_superpod} and driving the data center (DC) power demand. BCG~\cite{bcg_report_24} projects the DC energy (likewise, power) demand in the US to rise at a CAGR (cumulative annual growth rate) of $15$-$20\%$ \cite{semianalysis_report} year-on-year, potentially reaching $1$,$000$~TWh ($100+$~GW) by $2030$---two-thirds of the US residential consumption. Similar trends are seen in the EU~\cite{mckinsey_eu_ai_power}.

Sensing this surge in power demand, the risk of power-supply bottlenecks~\cite{bill_gates_blog, berkeley_lab_queue}, and mounting pressure to meet sustainability goals, hyperscalars like Microsoft, Google, and others have recently announced partnerships~\cite{ms_energy_deals, google_20B_deal, google_nuclear, ms_three_mile} with many renewable\footnote{We discuss challenges with nuclear energy in~\cref{sec:discussions}.} energy providers.

Unfortunately, simply adding more renewables to the utility grid may not suffice.
In particular, as power demand surges, grids struggle to keep up.
First, expanding power infrastructure---via new transmission lines, distribution systems, or energy storage---is capital-intensive~\cite{TnD_cost, fares2017trends}, often faces regulatory and logistical delays~\cite{bill_gates_blog}, and is especially difficult when renewable sources are located far from consumption hubs~\cite{china_curtailment, wind_far}.
As a result, a Berkeley Lab report~\cite{berkeley_lab_queue} highlights that by 2023, pending grid approvals for new power generation now exceed twice the installed capacity.
Second, even approved projects frequently face curtailment due to grid congestion, leaving clean power underutilized~\cite{eia_curtailment, uk_curtailment, china_curtailment}.
Thirds, long-distance transmission and distribution (T\&D) can incur significant power loss, inflating costs by over $50\%$\cite{fares2017trends, virtual_battery}.
For example, the EIA reports~\cite{eia_monthly_price} US enterprise power costs at $8.5$~\cent/KWh (Sep' $24$), while wind farms sell at $2.5$~\cent/KWh\cite{wind_ppa_2, wind_ppa_1}. 
Finally, much of the grid is aging---approaching or exceeding its 50-year design lifespan~\cite{bill_gates_blog, goldman_sachs_may24}---and may be forced to adopt short-term measures~\cite{evergy, dominion_energy} that compromise long-term sustainability goals of all stakeholders~\cite{energy_mix}.


Observing the gap, we ask the questions: \emph{Can we offer AI providers a useful lever for addressing power constraints and advancing sustainability goals?}
In this light, we propose \textit{AI Greenferencing} that shifts a significant share of AI inferencing to renewable energy farms, with a focus on wind farms due to their growing deployment~\cite{windcores, soluna, westfalenwind}. 
While these sites currently support digital television streaming and crypto-mining, LLM inferencing ($90\%$ of AI compute today~\cite{patel2024characterizing, amazon_inference}) presents a major opportunity.
Running AI at the power source eases grid stress, addressing interconnection queues, curtailment, T\&D loss, and sustainability.
AI Greenferencing is broadly a win-for-all: 
(1) users gain access to sustainable AI services; (2) AI providers unlock additional compute capacity, user reach, and revenue; (3) wind farms can monetize power locally; (4) power grids benefit from reduced load and more breathing room for expansion; and (5) co-located compute startups~\cite{soluna, windcores, westfalenwind} gain a high-value workload.

While conceptually simple, realizing AI Greenferencing requires navigating several practical questions:

\subheading{Enough wind power to make a dent?} We find that there already lies a significant wind capacity (only $100+$~MW deployments) of $640+$~GW ($150+$~GW) globally (likewise, in the US), as per the Global Energy Monitor data~\cite{wind_data_global_energy_monitor}, within $50$~ms fiber round-trip time (RTT) of Azure DCs. $77\%$ of this capacity is within $20$~ms RTT (does not significantly inflate AI inferencing latency demands with time-to-first-token or TTFT SLOs of up to few seconds) of Azure DCs~\cite{azure_dc_locations}. 

\subheading{AI compute too costly?} We show that it is economically feasible to deploy costly GPUs at wind farms with variable power. In particular, there exist multiple modular compute deployment choices~\cite{microsoft_mdc, nvidia_superpod, mdc_adv, micro_new_mega} which could be mass produced and shipped to wind farms. 
Our deployment right-sizing strategy recommends deploying AI compute at a lower-than-peak power generation.
This approach guarantees peak compute power most of the time with residual uncertainty that is handled by the software router leveraging cross-site complementarity~\cite{virtual_battery, skybox} of generation.
The hardware deployed on a wind farm incurs significantly lower power cost at the source $100\%$ of the time, which translates to a comparable or lower total cost of ownership (TCO) than in traditional DCs.
With this approach, we could deploy $6+$ million NVIDIA H$100$s ($\geq3\times$ NVIDIA's order book for $2024$~\cite{nvidia_orderbook_1, nvidia_orderbook_2}) in wind farms `today.' 

\subheading{Variations predictable?} We find that wind power albeit variable is highly predictable at 15-min timescales, thus assisting the scheduler in taking informed decisions on routing online.
Similarly, AI inferencing workload arrival is predictable allowing AI Greenferencing software to plan allocation ahead with power constraints.


\begin{figure}[t]
    \centering
    \includegraphics[width=0.85\columnwidth]{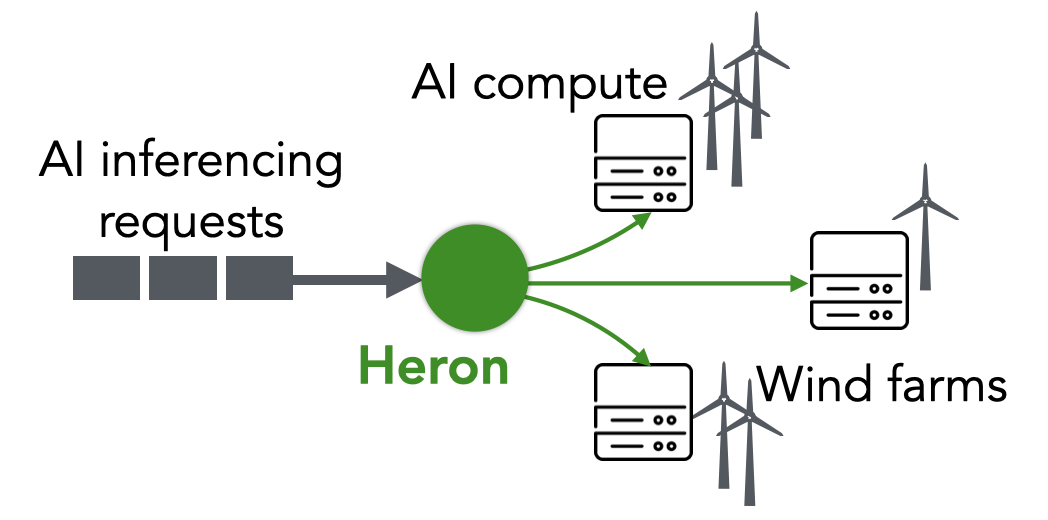}
    \caption{AI Greenferencing with \rname.}
    \label{fig:ai_greenferencing}
\end{figure}

\parab{\rname for AI Greenferencing.} When co-locating GPUs with wind sites, another challenge is to servce AI inferencing requests under variability of power and workload~(\cref{sec:Design}).
To realize this vision, we design \rname (shown in \cref{fig:ai_greenferencing}), a logically centralized cross-site router that intercepts inference requests near the AI provider's edge or datacenter, and routes them to wind-powered compute clusters.
\rname accounts for power availability, hardware constraints, network latency, and workload patterns.
It exploits both application-level knobs (e.g., tensor parallelism) and infrastructure-level controls (e.g., frequency scaling) while leveraging spatial complementarity across wind sites.
Hierarchically structured for multi-timescale decisions, \rname improves AI compute goodput by up to 80\% over a state-of-the-art energy-aware but variability-agnostic baseline scheduler~\cite{dynamollm}.


This paper makes the following contributions:
\begin{itemize}[left=0pt]
    \item We show there is significant wind capacity close to demand.
    \item We show how right-sizing the AI compute per wind site could largely \emph{mitigate the power uncertainty}.
    \item We show how low-cost power at its source could offset lost compute cycles due to the residual uncertainty.
    \item We show that both wind power generation and inferencing workload arrival are variable yet \textit{predictable}.
    \item We build and demonstrate \rname, a cross-site router for AI Greenferencing that leverages predictions, workload insights, application and infrastructure-level knobs, and wind generation complementarity across sites to \textit{address the residual uncertainty} (post right-sizing) in the system.
\end{itemize}



\section{Motivation \& Feasibility analyses}
\label{sec:feasibility}

\noindent AI compute is becoming increasingly power-dense~\cite{power_density_1, power_density_2, power_density_3}, with datacenter demand in the U.S. alone growing at a CAGR of 10--15\%~\cite{bcg_report_24}, soon rivaling residential consumption.
This accelerating demand risks overwhelm aging grid infrastructure and force short-term decisions~\cite{evergy, dominion_energy} that may undermine long-term sustainability goals.

Meanwhile, wind farms represent a massive, underutilized power opportunity.
As of June 2024, there is 447~GW of operating global wind capacity and over 2.4~TW in total projected capacity (including under-construction, pre-construction, and announced $100+$~MW sites)~\cite{wind_data_global_energy_monitor}.
However, much of this capacity remains idle---stuck in interconnection queues~\cite{berkeley_lab_queue} due to the high cost, complexity, and slow pace of grid expansion.
Even approved wind farms often face curtailment~\cite{china_curtailment, uk_curtailment, eia_curtailment} due to grid congestion.
And when wind power is delivered via the grid, transmission and distribution (T\&D) losses---often exceeding 50\%~\cite{fares2017trends, virtual_battery}---inflate costs significantly.

\parab{A case for AI greenferencing.}
This paper argues a disaggregated AI deployment framework across large wind farms that co-locates `AI' compute clusters with power sources.
Rather than bringing power to compute, AI greenferencing framework routes LLM inferencing  requests across geographically dispersed sites, adapting to workload characteristics, power availability, generation complementarity, and hardware capacity.
This model has the potential to unlock millions of GPUs for a sustainable AI infrastructure.

This section quantifies the opportunity and evaluates key feasibility dimensions:  
\S\ref{subsec:capacitywithinrange} assesses wind capacity near existing demand;  
\S\ref{sec:provisioning} analyzes the cost trade-offs of sacrificing compute cycles for cheaper power;
and \S\ref{subsec:router_feasibility} examines how to tame fluctuations in wind power and workload.

\begin{figure*}[tbh]
  \centering
  \subfigure[]{\label{fig:connectivity:global_map}
    \includegraphics[width=0.45\textwidth]{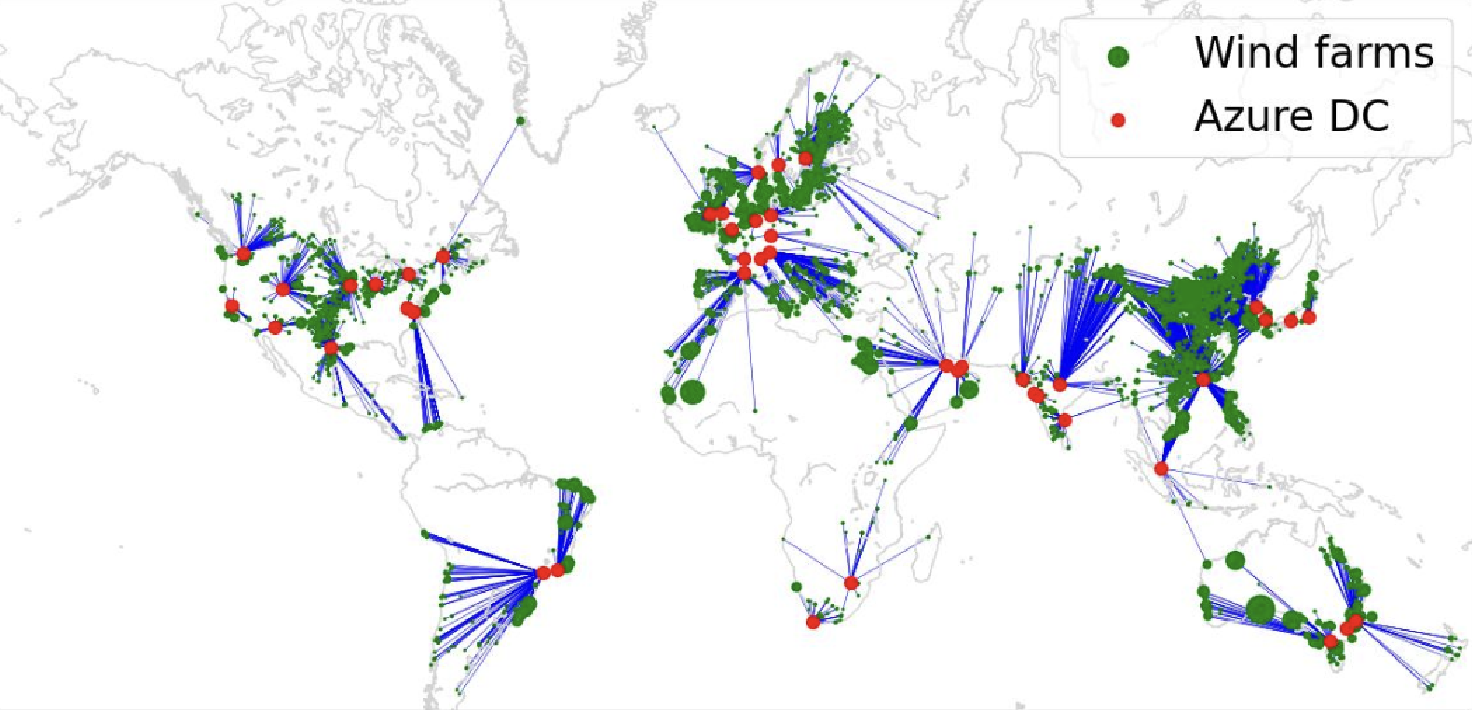}}
  \subfigure[]{\label{fig:connectivity:capacity}
    \includegraphics[width=0.45\textwidth]{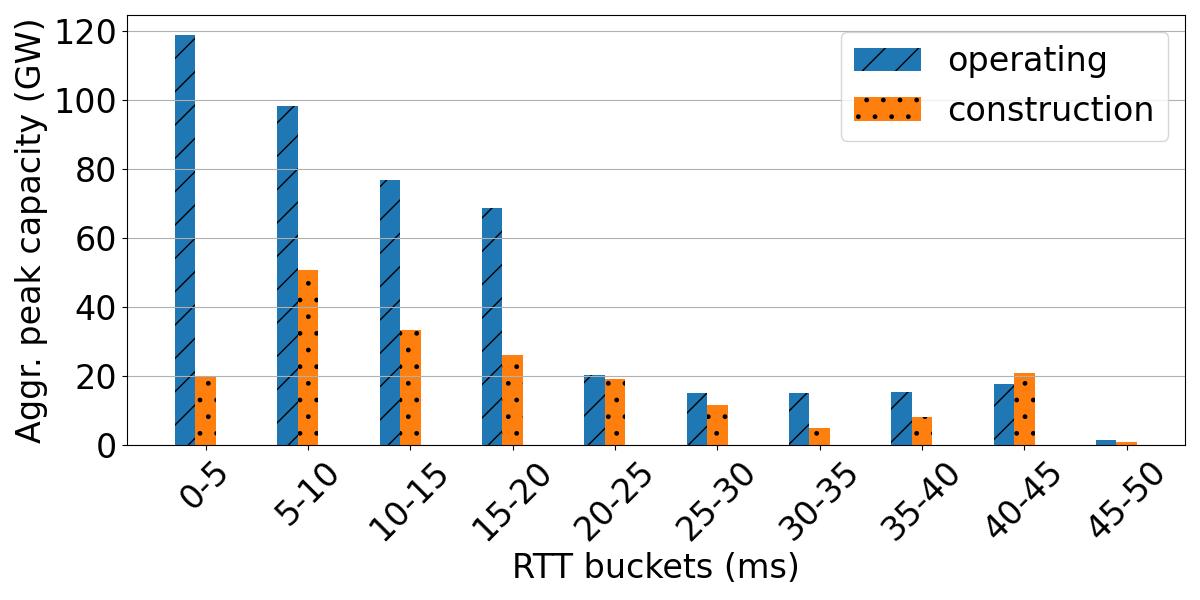}}
  \caption{(a) Wind capacity ($100+$~MW farms only) within a $50$~ms (emulated) fiber RTT of Azure locations as of June '$24$. 
  (b) A significant fraction of this peak capacity lies within $20$~ms fiber RTT.}
  \label{fig:connectivity}
\end{figure*}

\subsection{Significant Wind Capacity Reachable}
\label{subsec:capacitywithinrange}


\noindent Checking for wind farms within reasonable network distance of DCs (demand proxies) is crucial.
AI inference latency service level objectives (SLOs)~\cite{dynamollm} include Time To First Token (TTFT) and Time Between Tokens (TBT). 
TTFT ranges from a few $100$~ms to seconds, while TBT is lower\cite{distserve:osdi24}. 
Network latency affects primarily TTFT, not TBT (streaming tokens).

Fig.~\ref{fig:connectivity:global_map} shows wind farms ($\geq100$~MW, as of June '$24$~\cite{wind_data_global_energy_monitor}) within $50$~ms RTT (we account for fiber latency inflation~\cite{bozkurt2018dissecting}) of Azure DCs (public, $2023$~\cite{azure_dc_locations}). 
This map represents a hub-and-spoke model where AI requests reach Azure DCs before routing to wind farms. 
In real deployments, a cross-site router at the Azure network edge could further reduce this extra latency.
Fig.~\ref{fig:connectivity:capacity} shows much of this capacity is within $20$~ms RTT, inflating TTFT (Table.~$IV$ in~\cite{dynamollm}) by only $1$-$8\%$. 
Similar analysis with Google data centers~\cite{gcp_dc_locations} reveals comparable proximity (not shown).
Not all capacity in Fig.~\ref{fig:connectivity} is available for AI deployment though. 
Wind power is uncertain, and \S\ref{sec:provisioning} shows how right-sizing compute at wind sites ensures economic feasibility. 
Note that while operating capacity serves grids, consuming power at the source instead of via the grid should not, in principle, disrupt the power supply ecosystem.  

\greybox{\textbf{Takeaway:} Significant wind power capacity exists near data center hubs (i.e., AI demand) without violating latency SLOs.}

\subsection{Hardware Provisioning}
\label{sec:provisioning} 

\parab{Hardware deployment choices.} Microsoft has been deploying Modular Data Centers (MDCs)~\cite{microsoft_mdc} for many years. MDCs could be mass-produced and assembled on site, hence incurring lower costs of production~\cite{mdc_cost_1, mdc_cost_2, mdc_adv} than traditional DCs. Also, there are multiple startups~\cite{soluna, windcores} that have been deploying and running compute in wind farms (and even within turbines). NVIDIA recently announced SuperPODs~\cite{nvidia_superpod} that are modular clusters of $1$,$016$ GPUs (like, H$100$) with integrated cabling, cooling, network, etc. There also exist past research (Parasol~\cite{parasol}) on running compute co-located with renewables. Given these capabilities, we rather focus on a more important question---does it make sense economically to deploy AI compute in wind farms?

A major challenge in deploying GPUs at wind farms is power variability (\S\ref{subsec:router_feasibility}). Given high GPU costs~\cite{h100_cost}, we assess if variable but cheap power could justify the CAPEX (capital expenditure) investment. GPUs have a limited lifespan, and underutilization leads to wasted CAPEX. We compare OPEX (operational expenditure) vs. CAPEX to determine optimal provisioning, ensuring lower OPEX offsets lost CAPEX.  



\begin{figure}[t]
    \centering
    \includegraphics[width=\columnwidth]{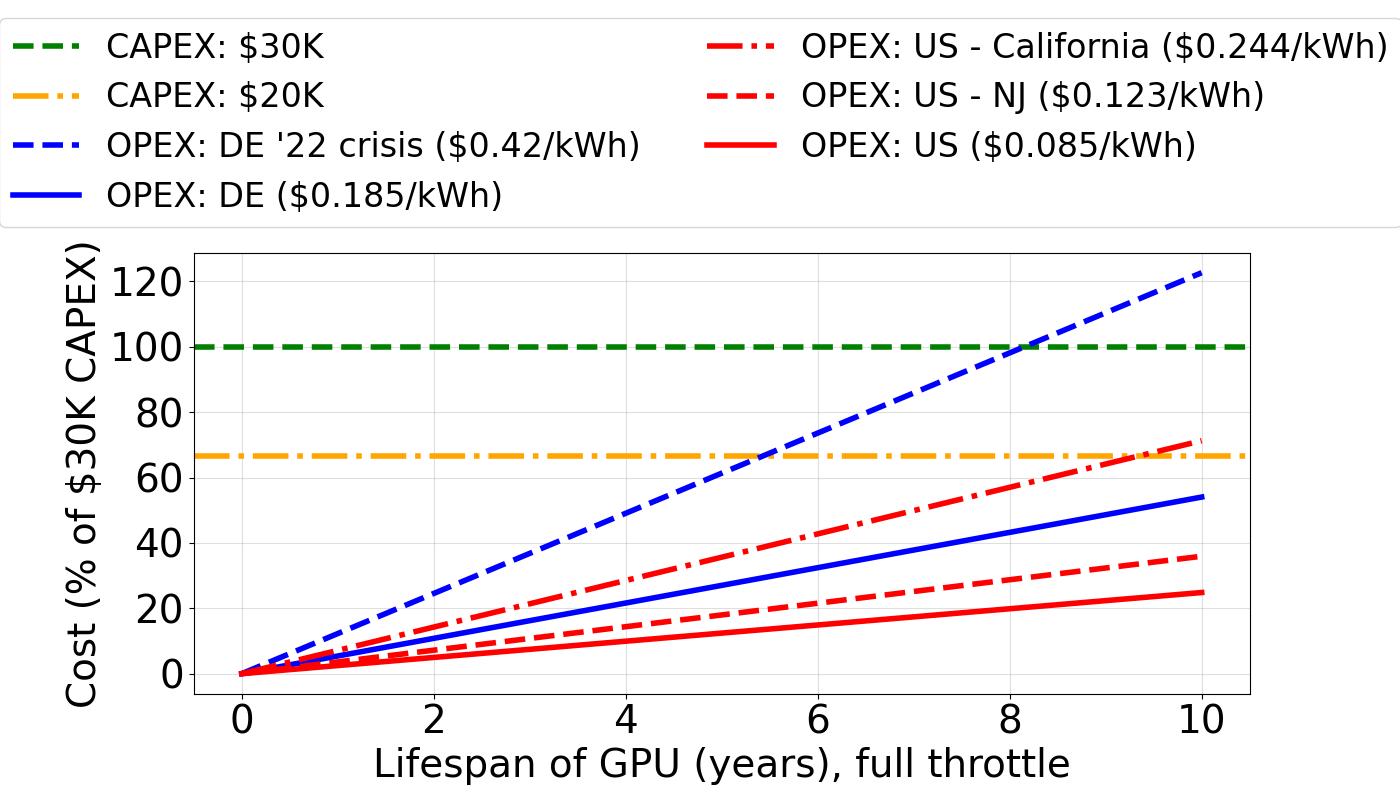}
    \caption{Cost of power over the lifetime of an NVIDIA H$100$ GPU could be a significant fraction of its CAPEX.}
    \label{fig:provisioning:cost_vs_time}
\end{figure}


\parab{OPEX versus CAPEX.} The total cost of ownership (TCO) of a DC is primarily driven by $4$ components~\cite{barroso2019datacenter}: ($1$) DC CAPEX (as DC depreciation), ($2$) DC OPEX, ($3$) Server CAPEX (as server depreciation), and ($4$) Server OPEX. MDCs are known~\cite{mdc_cost_1, mdc_cost_2, mdc_adv} to have a low cost of production due to economies of scale -- they could be mass produced. Furthermore, when deployed on otherwise unused land in a wind farm, the cost of land acquisition or rent should also be low~\cite{land_use_1, land_use_2} given the revenue opportunity this deployment brings to wind farms. Hence, we do not anticipate the DC CAPEX to be any higher for AI Greenferencing deployments than for traditional DCs. In addition, MDCs are easier to maintain given their standard design at scale. If wind farm employees/operators are trained to perform basic DC maintenance activities~\cite{dc_maintenance, hong2024self}, this could rather bring down the DC OPEX cost than what a traditional DC incurs. Also, compute at wind farms could naturally leverage air-based cooling, effectively reducing reliance on liquid cooling and, hence, reduce OPEX further. Hence, here we focus more on only the server CAPEX and OPEX aspects---the former is primarily driven by GPU prices, and the latter, driven significantly by the cost of power, could be significantly lower with AI Greenferencing style deployments.

Given GPU costs have largely increased in the last few years~\cite{h100_cost} with NVIDIA H$100$s priced at around $30K$~USD per GPU, it could be easy to believe that CAPEX should drive server deployment choices, as the cost of power is traditionally low. Here we show otherwise -- over the lifetime of a GPU, the OPEX (cost of power in our analysis) becomes a significant fraction of the TCO. Fig.~\ref{fig:provisioning:cost_vs_time} shows the CAPEX and OPEX components of an NVIDIA H$100$ GPU normalized by its publicly known price of $30K$~USD~\cite{h100_cost} (horizontal line at $y=100$). 
A second horizontal line (CAPEX of $20K$~USD) represents a plausible lower bound of the economies of scale for bulk buyers. All other lines show the total OPEX spent in buying power in the US~\cite{eia_monthly_price} and in Germany (DE)~\cite{germany_electricity_cost} anticipating peak GPU operations throughout. As we can see, in $5$ years, the OPEX in the US is $12.4\%$ ($18.6\%$) assuming $30K$~USD (likewise, $20K$~USD) CAPEX. In Germany these numbers are much higher---$27\%$ and $40.5\%$ respectively. In special cases, when compared against a $30K$~USD CAPEX, OPEX could be as high as $35.6\%$ in the US (California), and $61\%$ in Germany ($2022$ crisis~\cite{germany_crisis_2022}).


\greybox{\textbf{Takeaway:} OPEX could be a significant fraction of the cost of ownership of even high-end GPUs.}

\begin{figure}[t]
    \centering
    \includegraphics[width=\columnwidth]{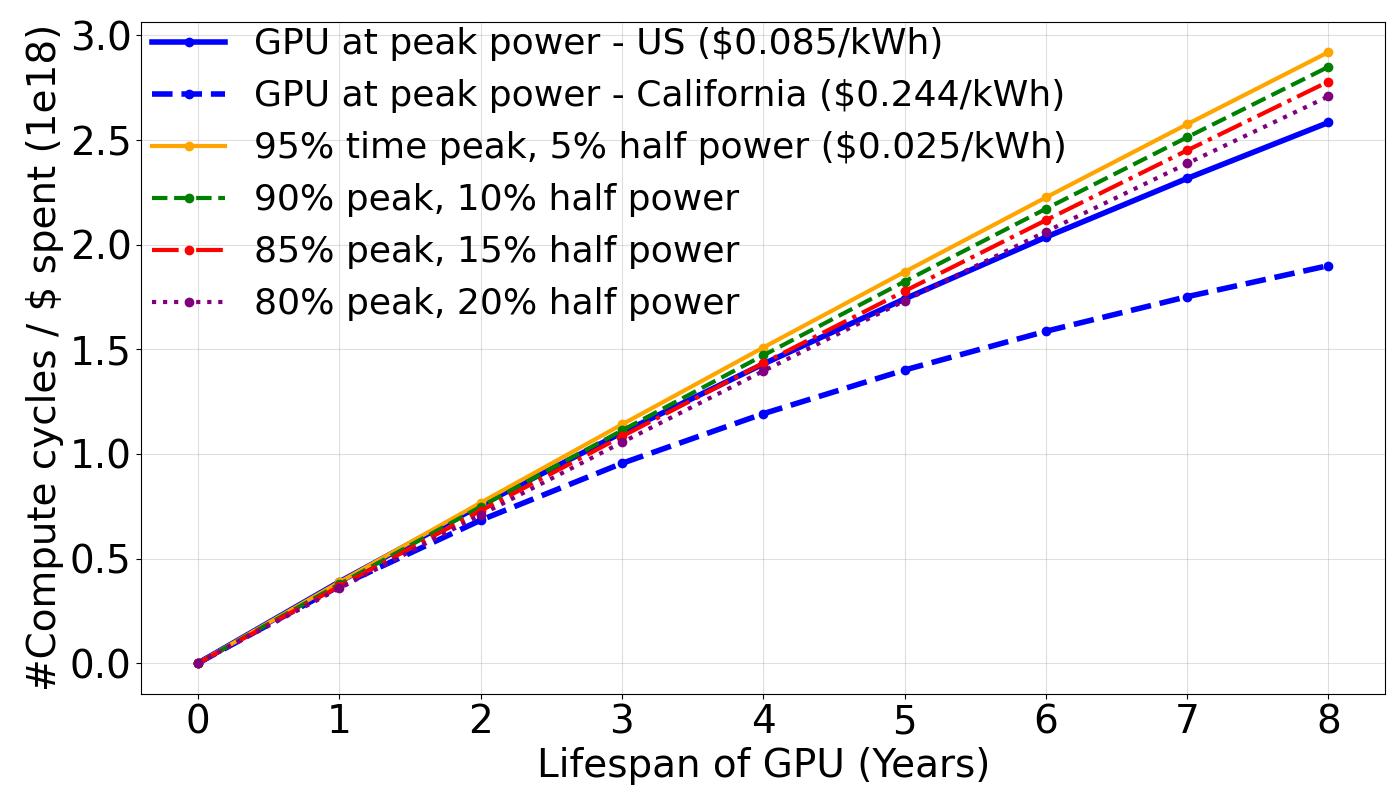}
    \caption{Comparable C/P at wind sites.}
    \label{fig:provisioning:percentile_vs_metric_US_total}
\end{figure}


\parab{Better C/P than in traditional DC.} As the cost of power consumption is a significant fraction of the TCO of AI compute, we argue that it could be a better choice to lose a few GPU cycles during its lifetime if the cost of power is significantly cheaper throughout. Let us see here how a GPU deployment in a wind farm could fare against one in a traditional DC in terms of capability/price (C/P = total compute cycles / dollars spent) where the denominator is the sum of CAPEX and OPEX. 

Fig.~\ref{fig:provisioning:percentile_vs_metric_US_total} plots C/P for a single GPU priced at $25K$~USD over time. 
Let's say that the GPU at its peak could sustain $\sim${}$10^{22}$~flops per year~\cite{h100_datasheet}. The cost of power for enterprise customers in the US is $0.085$~USD/kWh on average, with the cost in California being $0.244$~USD/kWh as of Sep'$24$ (EIA Report~\cite{eia_monthly_price}). On the other hand, the cost of wind power at the source, as in reported PPAs~\cite{wind_ppa_1, wind_ppa_2} (power purchase agreements), is significantly lower\footnote{As per industry experts, often an order of magnitude cheaper.}---$0.025$~USD/kWh in the major wind power producing regions in the US. This lower cost of power translates to significant OPEX reductions when AI compute is deployed at the source of power, that is the wind farms. Also, note that this lower cost is actually an upper bound for the cost of power at its source than via grid---curtailed power is anyway wasted and is virtually \textit{free}.

Provisioning for peak wind power at a farm wastes compute cycles. Instead, we provision at the $X^{th}$ percentile of generation, ensuring cheaper power $100\%$ of the time while minimizing under-utilization. Fig.~\ref{fig:provisioning:percentile_vs_metric_US_total} shows that this strategy achieves higher C/P than traditional DCs within years. At the $5^{th}$ percentile, parity is reached in $2$ years; at the $15^{th}$ \& $20^{th}$ (larger clusters), it takes $4$ \& $5$ years respectively.


\greybox{\textbf{Takeaway:} Right-sizing the AI compute provisioned in wind farms and consuming low cost power at source could effectively offset lost compute cycles.}


\begin{figure}[t]
    \centering
    \includegraphics[width=0.85\columnwidth]{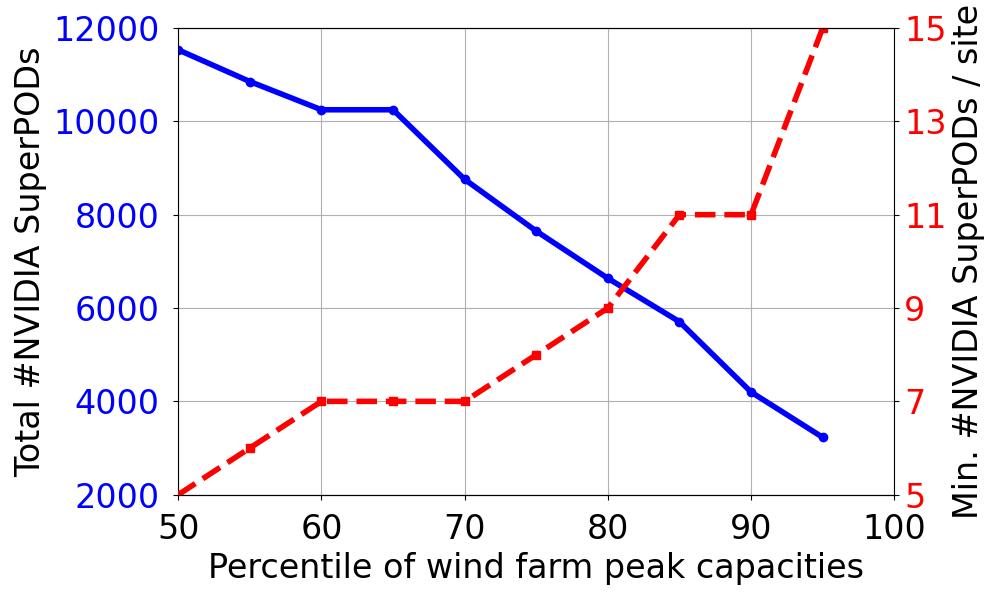}
    \caption{Deploying less compute ($20^{th}$ ptile of generation) at the larger farms ($x$-axis) translates to less fragmentation (right-$y$) but a large number of GPUs in aggregate (left-$y$).}
    \label{fig:provisioning:size_percentile_vs_hardware}
\end{figure}


\parab{High deployment appetite.} The key is to provision AI compute at a lower percentile of wind power generation at a site, ensuring comparable or higher C/P than traditional DCs despite occasional lost cycles. ELIA data~\cite{elia} shows that the $X^{th}$ percentile power is much lower than $X\%$ of peak capacity---e.g., ($5^{th}$~ptile, $1\%$), ($10$, $2$), ($15$, $4$), ($20$, $5$), etc. This approach translates to a fixed number of GPU clusters, scaling with farm size.


We now evaluate how deployment at the $20^{th}$ percentile of the peak farm capacity could look like globally, if we only pick the largest (in terms of peak generation capacity) $Y\%$ wind farms from the Global Energy Monitor dataset~\cite{wind_data_global_energy_monitor}. For this evaluation, we assume deployments will happen in multiples of NVIDIA's H$100$ SuperPODs, each hosting $1$,$016$ H$100$ GPUs and consuming $1.3$~MW peak power. As seen in Fig.~\ref{fig:provisioning:size_percentile_vs_hardware}, the aggregate number of NVIDIA SuperPODs that could be deployed at the largest $20\%$ ($x=80$, $250+$~MW sites) of the operating wind farms across the globe is $6$,$636$ -- this translates to  $6.7$~million H$100$ GPUs (more than $3\times$ the H$100$ order book of NVIDIA in $2024$~\cite{nvidia_orderbook_1, nvidia_orderbook_2}). Given these are all large wind farms, even the $20^{th}$ percentile ($5\%$) of peak capacity in the smallest such farm is a reasonable sized deployment of $9$ SuperPODs ($9K+$ H$100$ GPUs). In general as we pick a smaller percentile of peak power for compute right-sizing, larger the aggregate deployment but smaller the minimum deployment size (more fragmentation). This is a knob that AI inferencing providers should use based on factors like software router scalability, per-site maintenance overhead, physical security concerns, etc.

\greybox{\textbf{Takeaway:} Despite the right-sizing strategy, an estimated $6+$~million H$100$ GPUs could still be deployed at the larger wind farms today.}


\subsection{Taming Power \& Workload Uncertainty}
\label{subsec:router_feasibility}

\noindent AI Greenferencing, spanning multiple wind sites, needs to mitigate uncertainties at either end: ($1$) power generation at the wind sites and ($2$) inference workload arrival.
Below we explore the specific characteristics of wind power and the AI inferencing workload (next) to better understand how predictable these variations are---a characteristic that could be leveraged by the software router.


\parab{$2.3.1$ Wind Power Characteristics}



\parab{Predictability.} Wind power is known to be variable due to temporal changes in wind patterns and weather conditions. We accommodate this variability while provisioning AI compute at wind farms, as discussed in \S\ref{sec:provisioning}.
Albeit variable, wind power is predictable, and this characteristic could be leveraged by the AI Greenferencing software while routing workload across multiple wind RE sites. We observe that the mean autocorrelation score across $4$ different combinations of wind regions (Wallonia, Flanders) and power grids (Elia, Dso) at $15$~min granularity (Jan to Jul'$24$, ELIA dataset~\cite{elia}) with a lag of $1$ is $0.991$. Across $235$ wind farm sites in the EMHIRES dataset~\cite{emhires}, the mean (likewise, median) autocorrelation score with a lag of $1$ is $0.989$ ($0.99$) for $1$~year ($2018$-$19$) of generation data at a granularity of $1$~hour. These observations demonstrate strong predictability of wind power at different temporal granularities with time series or ML-based prediction models that could consume additional input features such as historical data, seasonality, local weather, turbine specification and orientation, farm layout, etc. Our custom-built AI-based energy prediction framework\footnote{Orthogonal to this work.} consumes diverse onsite and offsite inputs and integrates various open source models like TFT~\cite{google_tft} from Google and DeepMC~\cite{kumar2021micro} from Microsoft. The framework offers a relative improvement of $20\%$ in prediction errors compared to state-of-the-art predictors like TFT (only onsite data). Such orthogonal work helps treat wind power generation as an oracle in systems design.

\begin{figure}[t]
    \centering
    \includegraphics[width=\columnwidth]{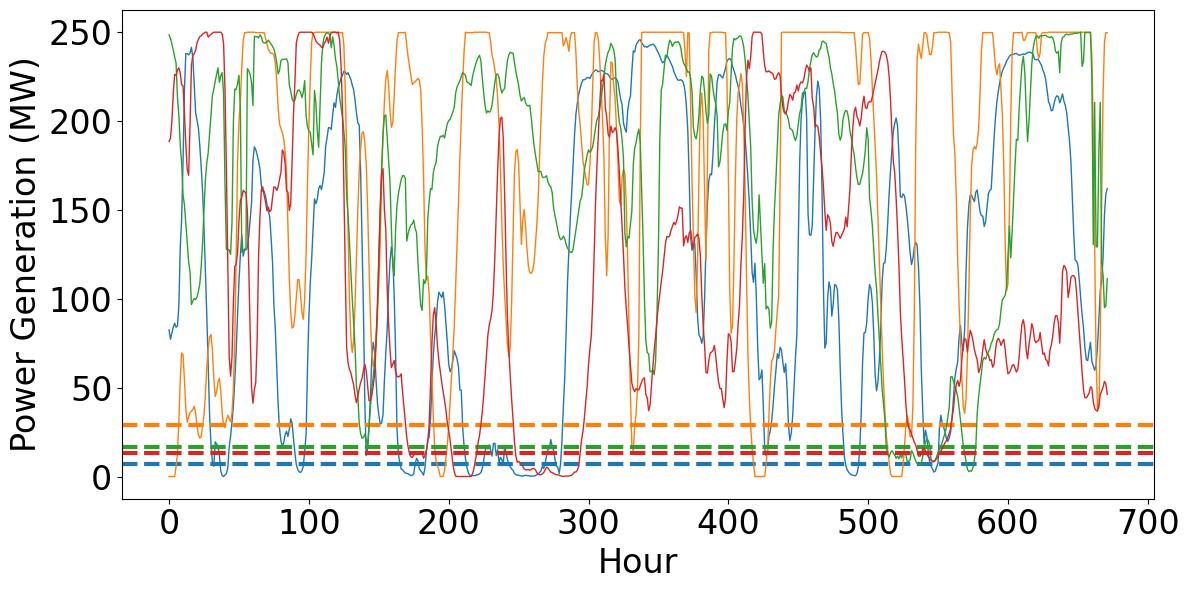}
    \caption{Scaled wind power generation across $4$ sites in Europe show complementarity. The dashed lines show their corresponding long-term (not just this window) $20th$~ptile generation.}
    \label{fig:wind_timeseries}
\end{figure}
\parab{Complementarity.} Wind power generation across sites could demonstrate complementarity, as already highlighted in past work, by leveraging different wind systems across geographies. While SkyBox~\cite{skybox} and Virtual Battery~\cite{virtual_battery} leverage complementarity to proactively migrate virtual machine workloads, \rname could use this characteristic of wind power to route AI inference requests around low generation. Note that each site is still constrained by the provisioned hardware, as we discuss in \S\ref{sec:provisioning}. \rname could route to a set of sites that offer enough complementarity such that the generation at all sites do not go down together. As there are usually many site choices within a range of $50$~ms fiber RTT (\S\ref{subsec:capacitywithinrange}), we could identify sets of cross-country sites that minimize the coefficient of variation (CoV) of aggregate generation. In the EMHIRES dataset, we found that combinations of $4$ wind sites (with assumed peak generation of $250$~MW following our discussion in \S\ref{sec:provisioning}; normalized to $1.0$ in EMHIRES) from multiple countries, all within a low network latency range, could reduce the CoV by $36\%$ (likewise, $32\%$) compared to a single site (likewise, $4$ different sites but within the same country). One such combination consists of $4$ wind farms from Iceland, Norway, Switzerland, and the UK, as seen in Fig.~\ref{fig:wind_timeseries} and offers a relatively low CV of $0.475$ long-term ($1$~year of data, EMHIRES). The dashed lines show the peak demand of the provisioned hardware ($20^{th}$ percentile of the power generation, long-term), as we discuss in \S\ref{sec:provisioning}. Note that the per-site generation is mostly higher than this peak demand and, also, very rarely the generation of all sites drop to the peak-demand thresholds at the same time.

\greybox{\textbf{Takeaway:} Wind power is predictable across time allowing software routing to plan-ahead. 
Spatial complementarity of wind power helps move around inference workload.}


\begin{figure}[H]
    \centering
    \includegraphics[width=0.85\columnwidth]{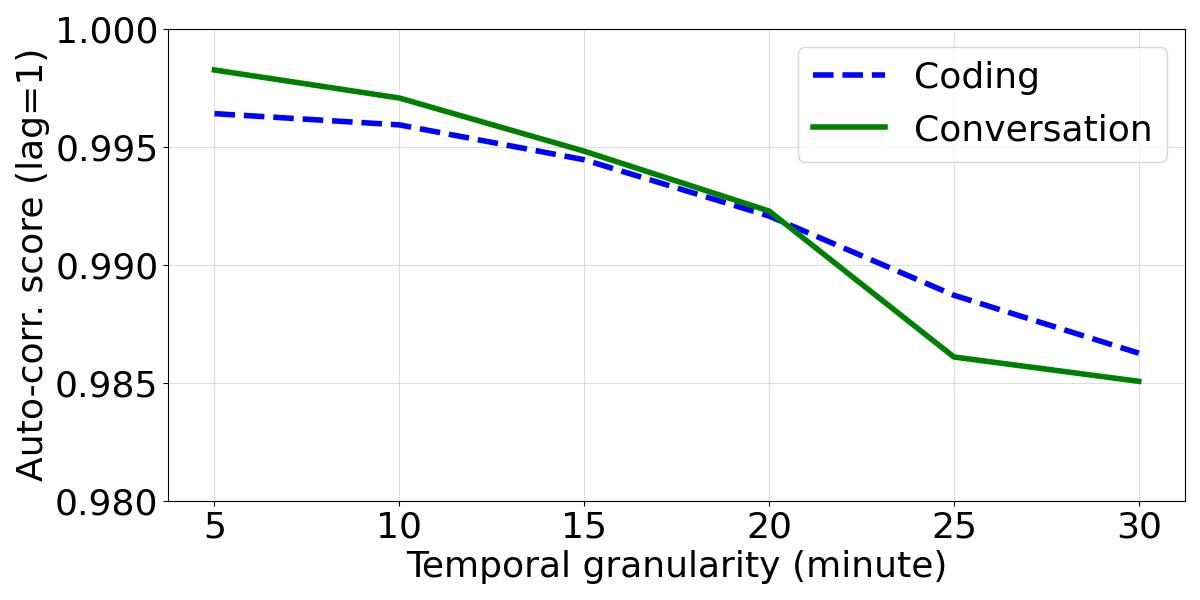}
    \caption{AI inferencing request arrival has high autocorrelation scores with a lag of $1$ for different temporal granularities.
    }
    \label{fig:autocorr:workload}
\end{figure}

\parab{$2.3.2$ Predictable AI Inferencing Workload}


\noindent We use real week-long \textit{coding} and \textit{conversation} production traces~\cite{azure_trace_dynamollm} from \cloudname and measure the variability and predictability. 
Both traces show clear diurnal and weekly patterns (Fig.~\ref{fig:trace}, right).
To understand the predictability of the workload we perform a simple time series analysis, as seen in Fig.~\ref{fig:autocorr:workload}, where we aggregate all request arrivals within a time window of $X$~min ($x$-axis; multiples of $5$) and compute the autocorrelation scores for a lag of $1$. As we can see, the scores are very close to $1.0$ in general. At $15$~min granularity, the correlation scores are higher than $0.994$ for both traces. This strong temporal correlation demonstrates that the workload arrival is highly predictable leveraging regression-based or machine learning based models with additional input features like time of the day, day of the week, etc. Workload predictability helps \rname make informed, intelligent decisions online.

\greybox{\textbf{Takeaway:} AI inferencing workload arrival is predictable thus allowing \rname intelligent planning ahead.}




\section{Design overview}
\label{sec:Design}

\noindent Motivated by the availability of wind power capacity close to demand and the predictable variability of wind generation, we now detail our cross-site router, \rname, that intelligently routes the AI inferencing requests across sites for good performance. \rname takes into account latency SLOs, wind generation variability, workload variability, workload characteristics, and application and infrastructure-level capabilities while routing inferencing requests. This section outlines key design challenges and opportunities, while the next delves into \rname's design components.




\subsection{Challenges}
\label{subsec:sys_design_challenges}

\begin{figure}[t]
    \centering
    \includegraphics[width=\columnwidth]{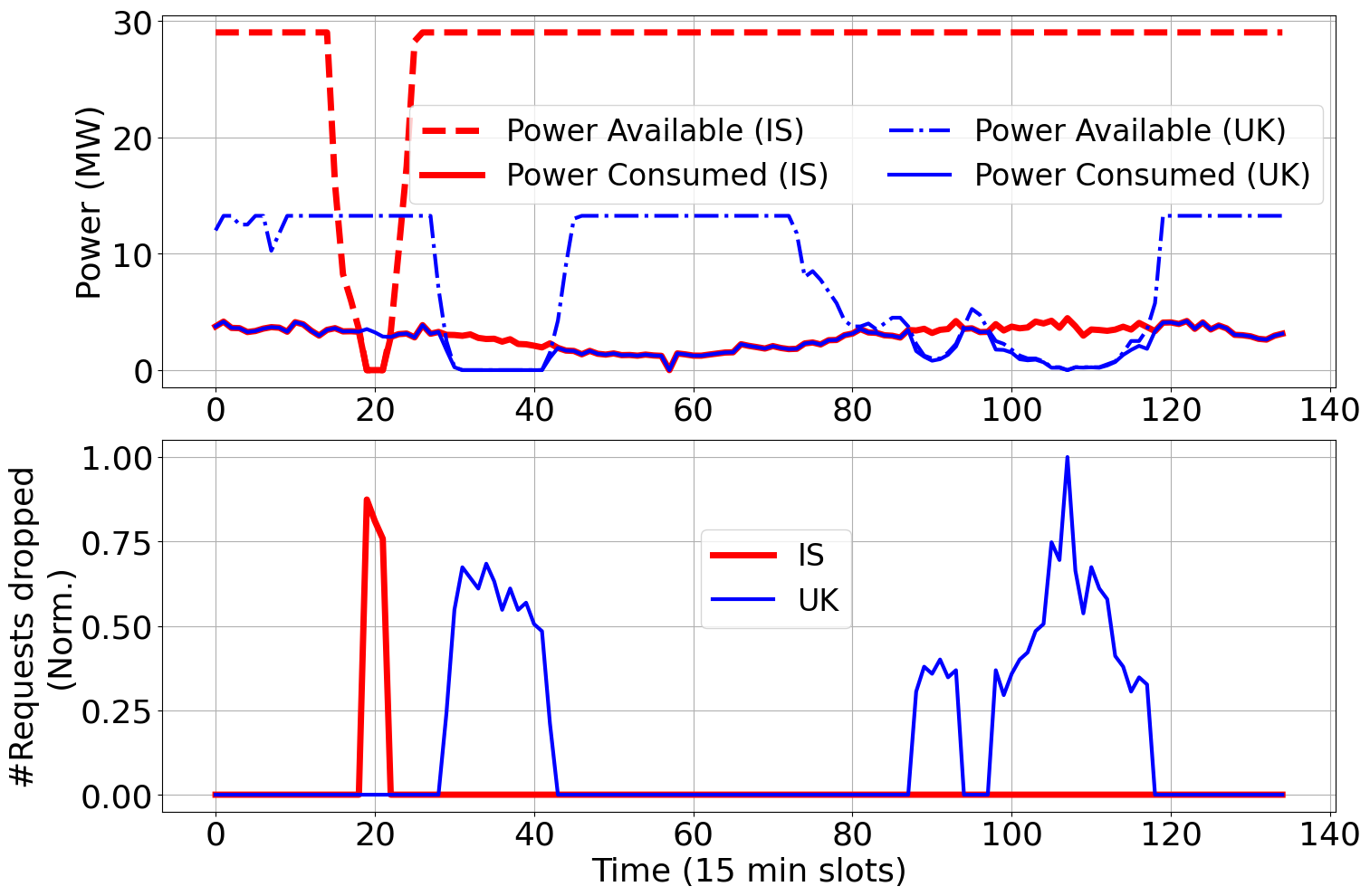}
    \caption{DynamoLLM needs to drop requests in the absence of cross-site intelligence that leverage cross-site complementarity.}
    \label{fig:dynamo:slo}
\end{figure}
\parab{(C$1$) Residual power uncertainty.} Right-sizing the compute deployment at each wind-site (see \S\ref{sec:provisioning}) significantly reduces the power uncertainty. Nevertheless, deploying compute, say, at the $20^{th}$ percentile of peak generation at a site still translates to lower-than-peak power $20\%$ of the time. A naive routing strategy could transiently bring more workload than a site can handle, given the power constraint, leading to request drops. Fig.~\ref{fig:dynamo:slo} demonstrates this challenge. The AI inferencing workload (real trace, discussed later in \S\ref{sec:results}) is routed across $2$ wind farm sites in the UK (compute provisioned for peak demand of $13.25$~MW) and Iceland (likewise, $29$~MW) following a round-robin strategy. Each site runs a DynamoLLM scheduler~\cite{dynamollm} (our baseline; discussed further in \S\ref{sec:related_work}) that aims to minimize power/energy consumption (assumes infinite hardware and power availability as in a traditional DC). We see that at times between $285$-$315$~min for the Iceland site and $435$-$630$, $1320$-$1395$, and $1470$-$1755$~min for the UK site, the available power drops below the provisioned peak and also the minimum ask of the assigned workload thus resulting in significant request drops (bottom plot, normalized by the peak drop). 

\parab{(C$2$) Complex workload characteristics.} In this work, we focus on large language model (LLM) inferencing. 
Different LLM use cases could have diverse input and output lengths, and latency SLOs, and could be run with different tensor parallelism (TP), GPU frequency, and load (requests per second or RPS) configurations. The configuration on which it runs decides the fate of the request---the power it consumes and the end-to-end (E$2$E) latency it incurs. In a cross-site dynamic-power and compute-constrained setting, this online matching problem poses a challenge.

\parab{(C$3$) Knobs with varying overheads.} The TP and GPU frequency knobs discussed above have overheads to different degrees. While changing GPU frequencies take milliseconds via \texttt{nvidia-smi}, changing TP degree could take a few seconds to minutes due to the multiple steps involved---waiting for existing requests to complete, re-sharding the model weights, transferring them to the GPUs involved, and synchronizing the GPUs. Tuning these knobs online in a cross-site setting, where each site could also have power variability, keeping in mind their varying overheads, needs careful planning.

\parab{(C$4$) Latency versus power trade-off.} Given the power and compute constraint, one could run the AI Greenferencing system to minimize either the aggregate request latency or the aggregate power consumption. We discuss in \S\ref{sec:results} how one could tradeoff latency for power and vice versa. While we leave it to the AI operator to decide on the operating point, AI Greenferencing design needs to be aware of both choices.

\subsection{Key Ideas}
\label{sec:overview:key}

\noindent \rname addresses the above challenges with the following key ideas:

\parab{(K$1$) Spatial power complementarity.} \textit{All} request drops in Fig.~\ref{fig:dynamo:slo} (C$1$) could have been avoided by routing around `bad' power conditions as the combination of sites offered full complementarity during the time window. \rname leverages this spatial complementarity to mitigate the residual temporal uncertainty, improving goodput over baselines (\S\ref{subsec:planner-L}). It uses a site-wise power oracle (\S\ref{subsec:router_feasibility}) while routing workloads.

\parab{(K$2$) Workload characteristics to lookup table.} To address C$2$, we ran a deep power profiling benchmark building on top of past work~\cite{dynamollm} but for more recent Llama $3.1$ $70$~B model. We classify inferencing requests based on their input and output lengths, quantify plausible latency SLOs, and quantify how different degrees of TP and frequencies could run the same request class (at different levels of load or RPS) at different power and latencies. The profiling exercise helps us obtain $2$ useful functions (lookup tables): \textit{e$2$e($c$, $f$, $t$, $l$)} that gives the end-to-end latency and \textit{power($c$, $f$, $t$, $l$)} that gives the peak power consumed by the $c$-th class of requests at load $l$ run with TP $t$ and GPU frequency $f$. The lookup table contains approximately $2$,$000$ rows. In a dynamic setting with available power being lower than the peak at times, it is important to leverage these tuning knobs well to remain within the resource constraints. We discuss profiling results in \S\ref{subsec:results:profiling} for both traces albeit profiling could be agnostic to use cases (not models). 


\parab{(K$3$) \rname's hierarchical design.} To tackle the overhead of different configuration knobs (C$3$), \rname is hierarchically designed---one component runs at a coarser temporal granularity deciding the TP degree of GPU instances, frequencies and load across all sites, while another runs more frequently reconfiguring only the frequencies and load. Such an architecture helps \rname adapt to differences in changing such knobs as mentioned in challenge C3, while still optimizing for real-time conditions. \rname minimizes the number of TP and frequency reconfigurations. This goes well with the fact that both power and workload changes are gradual too (see \S\ref{subsec:router_feasibility}). \rname also adopts DynamoLLM~\cite{dynamollm} optimizations (using inter-GPU NVLink and transferring weights between GPUs in the background) toward minimizing re-sharding overheads during TP reconfigurations. The hierarchical decision making effectively addresses the problem of mispredictions at coarser timescales.

\parab{(K$4$) \rname's different objectives.} \rname components could run with either a minimum latency or a minimum power consumption objective (C$4$). The different components work in tandem with the same objective, except a heuristic that aims to pack requests better toward improving the response latencies. Our results compare both these objectives and quantifies the tradeoff.

\begin{figure}[t]
    \centering
    \includegraphics[width=\columnwidth]{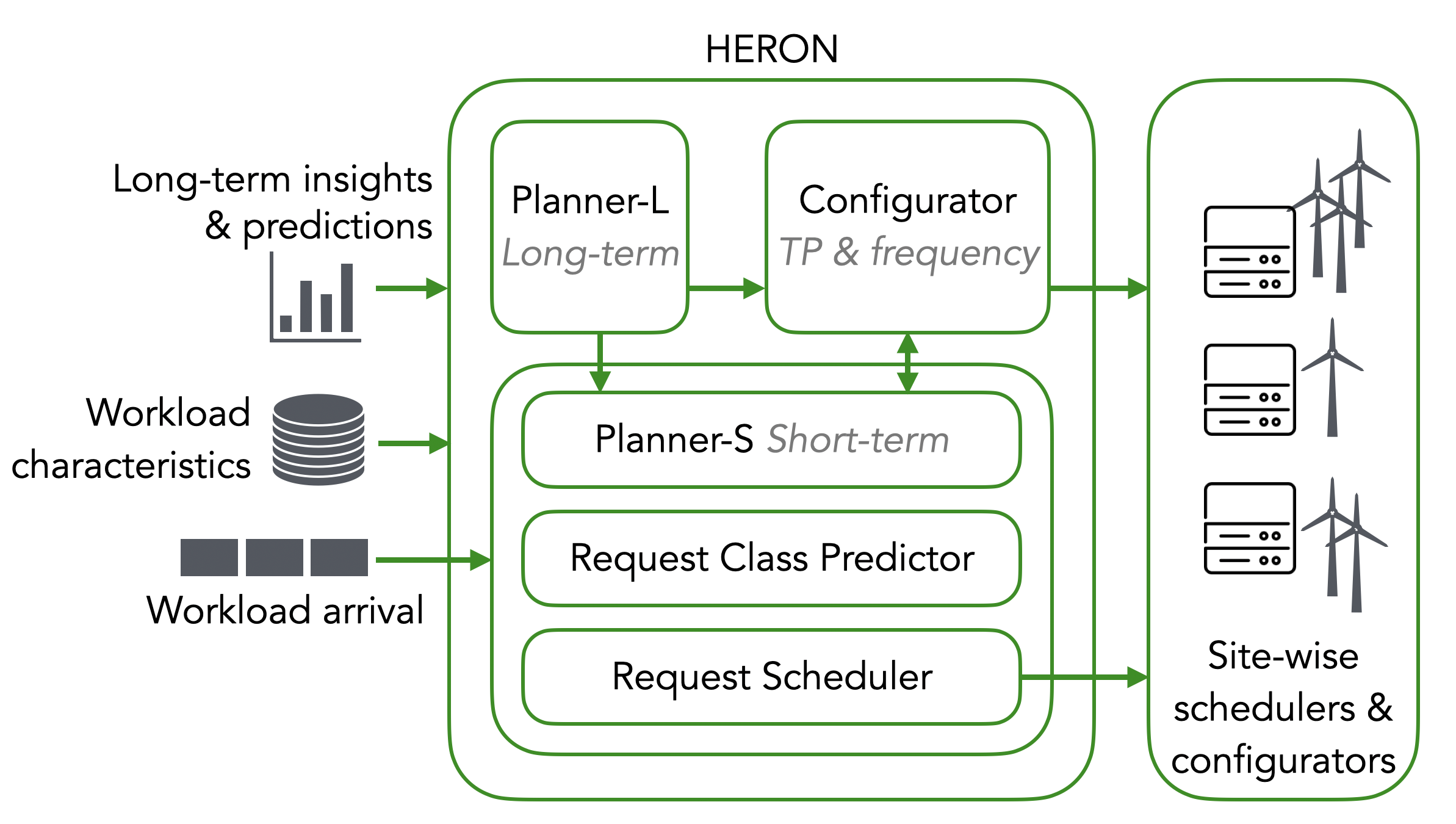}
    \caption{AI Greenferencing architecture.} 
    \label{fig:architecture}
\end{figure}

\section{\rname design}
\noindent \rname cross-site request router consists of multiple components as shown in Fig.~\ref{fig:architecture}. The goal of \rname is to determine the number of GPUs, TP degree, GPU frequency, and load at each renewable site for different request classes. Note that inference workloads could be split into multiple classes~\cite{dynamollm} based on input and output length, with each class having different power demands and latency SLOs (discussed later in \S\ref{subsec:results:profiling}). 
We decompose \rname into three key components that run at different timescales: \texttt{Planner-L}, \texttt{Planner-S}, and \texttt{Request Scheduler} with a \texttt{packing heuristic}. \texttt{Planner-L} runs at a coarser timescale (e.g., $15$~min) and assigns the TP degree, frequency, and load. \texttt{Planner-S} runs every few seconds and updates the frequency and load based on near-real-time power and workload conditions while sticking to the TP assignments done by \texttt{Planner-L}. A real-time \texttt{packing heuristic} runs with a \texttt{Request Scheduler} to further handle the changing conditions. We detail them below.


\begin{table}{}
\caption{Notations: ILP for Planner-L and Planner-S}
\label{tab:algo1:notation}
{
\small
\begin{tabular}{|p{0.27\columnwidth}|p{0.65\columnwidth}|}
\hline
\textbf{Notation} & \textbf{Description}\\
\hline
$C, T$ & Set of workload classes and TP degrees\\
\hline
$F, L$ & Set of all frequencies and loads (requests/s)\\
\hline
$S$ & Set of wind farm sites\\
\hline
$R_{L}$ & Limit on reconfigurations\\
\hline
$N_{s}, P_{s}$ & Number of GPUs (right-sizing, \S\ref{sec:provisioning}) and power at site $s$ (from real traces)\\
\hline
$Load_{c}$ & Peak load for class $c$ (from prod. trace)\\
\hline
$TP\_GPU(t)$ & Number of GPUs for TP degree $t$\\
\hline
$e2e(c,f,t,l)$ & End-to-end latency and power for class $c$,\\
$power(c,f,t,l)$ & frequency $f$, TP $t$, load $l$ (from profiling exercise -- lookup table)\\
\hline
$Old_{c,f,t,s,l}$ & Number of GPUs in previous iteration for a \{$c,f,t,s,l$\}\\
\hline
$GPU_{s,c,t}$ & Number of GPUs assigned to \{$s,c,t$\}\\
\hline
$X_{c,f,t,s,l}$ & Number of instances for a \{$c,f,t,s,l$\} \\
 \hline
$Y_{c,f,t,s,l}$ & Binary flag set if at least one instance assigned for a \{$c,f,t,s,l$\}\\ 
\hline
$R_{c,f,t,s,l}$ & Number of reconfigurations for a \{$c,f,t,s,l$\}\\
\hline
$Z_{c,f,t,s,l}$ & Number of GPUs calculated by Planner-S\\
\hline
\end{tabular}
}
\end{table}

\begin{figure}[t]
{\small
\fbox{
  \begin{minipage}{0.47\textwidth}
    \textbf{ILP Variable:} $X_{c,f,t,s,l}$\\ 
    \textbf{Objective:} Minimize $\displaystyle\sum_{c \in C, f \in F}\displaystyle\sum_{t \in T, s \in S, l \in L} X_{c,f,t,s,l} \cdot e2e(c,f,t,l)$\\
    \textbf{Constraints:}\\
    ($1$) GPU per site within limit: \\$\forall s \in S, \displaystyle\sum_{c \in C, f \in F, t \in T, l \in L} X_{c,f,t,s,l} \cdot TP\_GPU(t) \leq N_{s}$ \hfill\\
    ($2$) Power per site within limit: \\$\forall s \in S, \displaystyle\sum_{c \in C, f \in F, t \in T, l \in L} X_{c,f,t,s,l} \cdot power(c,f,t,l) \leq P_{s}$ \hfill\\
    ($3$) Enough serving capacity for each class: \\$\forall c \in C, \displaystyle\sum_{s \in S, f \in F, t \in T, l \in L} X_{c,f,t,s,l} \cdot l \geq Load_{c}$ \hfill\\
    ($4$) One configuration of f and l: \\$\forall s \in S, c \in C, t \in T, \displaystyle\sum_{f \in F, l \in L} Y_{c,f,t,s,l} \leq 1$ \hfill\\
    ($5$) Expressing Y in terms of X: \\$\forall s \in S, c \in C, t \in T, f \in F, l \in L, 1 \geq Y_{c,f,t,s,l} \geq \frac{X_{c,f,t,s,l}}{sum(N_S)} $ \hfill\\
    ($6$) Number of reconfigurations are bounded: \\
    $\displaystyle\sum_{c \in C, f \in F}\displaystyle\sum_{t \in T, s \in S, l \in L} R_{c,f,t,s,l}  \leq R_{L}$ \hfill\\
    (7) Expressing $R_{c,f,t,s,l}$, $\forall s \in S, c \in C, t \in T, f \in F, l \in L$, \\
    $R_{c,f,t,s,l} \geq Old_{c,f,t,s,l} - X_{c,f,t,s,l}, R_{c,f,t,s,l} \geq X_{c,f,t,s,l} - Old_{c,f,t,s,l}$\hfill
  \end{minipage}
}
}
\caption{Formulation: ILP for Planner-L}
\protect\label{fig:plannerl:formulation}
\end{figure}
\parab{Planner-L.} Both workload arrival and wind power generation per site are highly predictable at $15$~min granularity (see \S\ref{sec:feasibility}) and these predictions, alongside inferencing profiled data (a lookup table), are leveraged by a long-term ($15$~min) lookahead \texttt{Planner-L} ILP (integer linear program). Based on the predictions, \texttt{Planner-L} comes up with a plan to assign the number of instances (of GPUs) for the next $15$~min window for the peak load of each class of inferencing requests (based on input and output length) at each site. 
In addition, it also assigns a single frequency and load for a TP degree for an individual class at a specific site. We applied this constraint to scale the ILP without affecting optimality. The values for classes and frequencies are detailed in \S\ref{subsec:results:profiling}.

\texttt{Planner-L} bounds TP reconfigurations to further aid the strategies for hiding reconfiguration overhead (discussed in \S\ref{sec:overview:key}). 
\texttt{Planner-L} outputs TP and frequency reconfigurations needed, the load (RPS) per GPU instance, and also the fraction of inferencing requests per class that should be routed to individual instances across the multiple sites. Table.~\ref{tab:algo1:notation} and Fig.~\ref{fig:plannerl:formulation} detail the notation and formulation of the ILP -- it takes as input the workload classes, TP degrees, GPU operating frequencies, load levels (RPS within latency SLO bounds) per class per instance, number of GPUs and power available per site in the next time slot, the peak load per class of requests, and previous configuration, and gives as output the optimal number of instances that minimizes end-to-end latency of inferencing. The ILP minimizes the aggregate end-to-end latency for the workload (thus not compromising on the user experience) making sure that the configuration respects the ($1$) hardware and ($2$) power availability constraints per site, ($3$) each class of requests get enough serving capacity that could satisfy the peak load, ($4$,$5$) a TP instance serving a class of requests get the same frequency assigned for the same load, and ($6$, $7$) the number of reconfigurations needed per time slot is bounded.

\parab{Configurator.} The \texttt{Configurator} is in charge of updating the site-specific configurators of the necessary changes in TP and operating frequencies of different clusters of GPUs. It restrains \texttt{Planner-S} from considering clusters with pending reconfigurations for workload placement.


\begin{figure}[tbh!]
{\small
\fbox{
  \begin{minipage}{0.97\columnwidth}
    \textbf{ILP Variable:} $Z_{c,f,t,s,l}$\\ 
    \textbf{Objective:} Minimize \\ $\displaystyle\sum_{s \in S, c \in C, t \in T} \displaystyle\sum_{f \in F, l \in L} Z_{c,f,t,s,l}  \cdot e2e(c,f,t,l)$\\
    \textbf{Constraints:}\\
    (1) Power per site within limit: \\ $\forall s \in S, \displaystyle\sum_{c, f, t, l} Z_{c,f,t,s,l} \cdot power(c,f,t,l) \leq P_{s}$ \hfill\\
    (2) Enough serving capacity for each class: \\
    $\forall c \in C, \displaystyle\sum_{s \in S, f \in F, t \in T, l \in L} Z_{c,f,t,s,l} \cdot l \geq Load_{c}$ \hfill\\
    (3) Limiting number of GPUs per config:
    \\$\forall s \in S, c \in C, t \in T, \displaystyle\sum_{l \in L, f \in F} Z_{c,f,t,s,l} \leq GPU_{s,c,t} $ \hfill
  \end{minipage}
}
}
\caption{Formulation: ILP for Planner-S}
\protect\label{fig:algo2:router}
\end{figure}

\parab{Planner-S.} While \texttt{Planner-L} runs every $15$~min and reconfigures TP, operating frequency, and load per GPU instance, we run a lightweight \texttt{Planner-S} every few seconds to tens of seconds (bounded by its scalability as the number of sites grow; see \S\ref{subsec:eval:all_components}) that reconfigures only the frequencies and the load for a configuration, as TP changes are heavier. 
Running \texttt{Planner-S} at finer timescales has $2$ major benefits:  ($1$) it could tackle near-real-time power and workload arrival uncertainties and ($2$) it could unleash performance opportunity by running requests at higher than \texttt{Planner-L} assigned GPU frequencies as the available power goes up, thus improving TTFT and TBT latencies.



\texttt{Planner-S} is also an ILP (notations in Table.~\ref{tab:algo1:notation} and formulation in Fig.~\ref{fig:algo2:router} that recalculates optimal frequency and load assignments for each configuration. Note that \texttt{Planner-S} has the same optimization goal as \texttt{Planner-L} (minimizes the end-to-end latency) but only 
reconfigures frequencies (significantly faster to do with minimal overhead) while working with TP calculated by \texttt{Planner-L}. The ILP variable $Z_{c,f,t,s,l}$ indicates number of instances assigned for each config (tuple of site, class and TP degree) and frequency and load. Note, only frequency and load are reassigned for individual configs. The constraints are: ($1$) power constraints are met, ($2$) the serving capacity is enough for the peak load, ($3$) fits in the same GPU budget assigned by Planner-L.

Note that \texttt{Planner-L} and \texttt{Planner-S} formulations could be trivially changed to minimize power consumption.

\parab{Request scheduler.} This component is in charge of dispatching the inference requests, as they arrive in the arrival queue, across multiple sites. It consumes the output of \texttt{Planner-S} to learn the weights with which it performs a weighted round-robin (WRR) routing for each class (input/output length) of requests across clusters of GPUs assigned to individual classes across sites. To assign classes to requests, it uses a \texttt{Request Class Predictor} that takes as inputs the request and the model, predicts the output length and, hence, the class. Our predictor consists of an Albert/DistilBert/Longformer~\cite{lan2020albertlitebertselfsupervised, sanh2020distilbertdistilledversionbert, beltagy2020longformer} layer for input embeddings followed by a regressor, and could predict the output length bucket (small/medium/long) with an accuracy of $99.95\%+$ (predicting buckets is easier than predicting absolute output lengths). Hence, following the footsteps of DynamoLLM~\cite{dynamollm}, we treat output length as an oracle in our experiments. The \texttt{Request Scheduler} also has a \texttt{packing heuristic} that we describe next. 

\parab{Packing heuristic.} If the \texttt{Request Scheduler} observes online that the load for a request class for a GPU instance is lower than expected, it moves a fraction of the smaller requests (e.g., LS) to this class (e.g., LM) for better performance. It starts with the larger requests, reassigns, and move to the smaller ones, freeing up resources at each step.


\parab{Meta choices.} ($1$) \rname can be deployed at the AI provider network ingress or near a DC to intercept workloads. ($2$) We exclude full-fledged DCs and on-site batteries/generators to stress-test \rname's ability to handle residual uncertainties in power and workload arrival. ($3$) While \rname supports any site-local scheduler, we use a FIFO round-robin scheduler for GPU assignment in this work.

 
\begin{figure*}[tbh]
    \centering
    \begin{minipage}{\textwidth}
        \centering
        \includegraphics[width=0.33\linewidth]{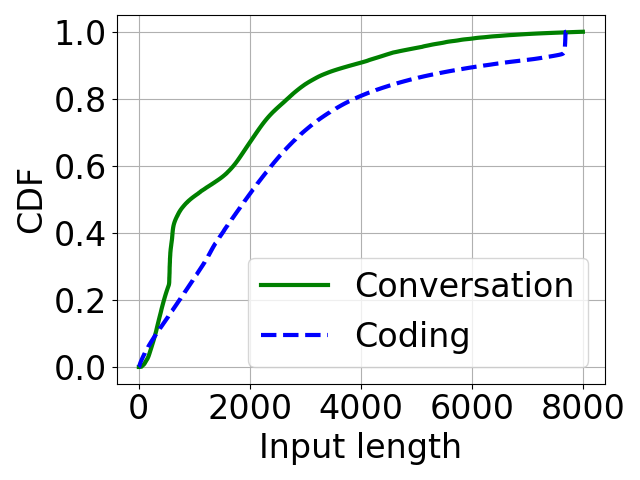}
        \hfill
        \includegraphics[width=0.33\linewidth]{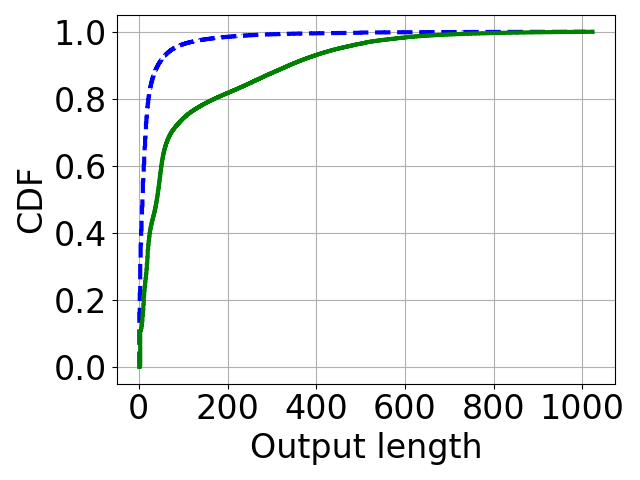}
        \hfill
        \includegraphics[width=0.33\linewidth]{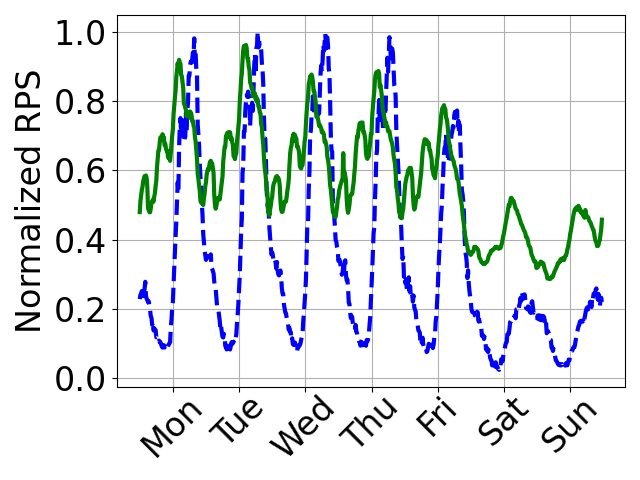}
    \caption{The input lengths (left), output length (middle), and arrival patterns (right) vary significantly across the $2$ traces and even for the same trace.}
    \label{fig:trace}
    \end{minipage}
\end{figure*}

\section{\rname Evaluation}
\label{sec:results}

\noindent We explore the following key questions (alongside others): \textbf{\S\ref{subsec:results:profiling}:} How do the power consumption and latency vary when workloads are run with different combinations of TP, frequency, and load? \textbf{\S\ref{subsec:planner-L}:} How does having cross-site global view help improve the offered goodput under uncertainties? \textbf{\S\ref{subsec:eval:all_components}:} Does a hierarchical \rname design unleash performance?

\parab{Data.} For the experiments below, we use the following data:

\parab{}($1$) $1$-week of LLM inferencing production trace for $2$ use cases, \textit{coding} and \textit{conversation}, on \cloudname~\cite{azure_trace_dynamollm}.
Fig.~\ref{fig:trace} shows the characteristic differences between the \textit{conversation} and \textit{coding} production traces on Azure. The inputs (left figure) vary between $1$ and $\sim${}$8K$ tokens, with \textit{coding} inputs being $\sim${}$2${}$\times$ larger than \textit{conversation} at the median. The outputs (middle figure) lie within $\sim${}$1K$ tokens with \textit{conversation} having $\sim${}$6${}$\times$ larger outputs at the $95^{th}$ percentile.

\parab{}($2$) Hourly power generation from $4$ wind farms in Iceland, Norway, Switzerland, and the UK (with assumed peak generation of $250$~MW~\footnote{We deploy compute at these larger sites to avoid significant fragmentation.} following our discussion in \S\ref{sec:provisioning}; normalized to $1.0$ in EMHIRES) for $672$ hours (see Fig.~\ref{fig:wind_timeseries}) in $2015$ from EMHIRES~\cite{emhires}. For our experiments, we scale the time dimension by $0.25$, thus translating it to $1$~week of power generation data at $15$~min granularity. We could not use the ELIA data (see \S$2.3.2$) as it covers only $2$ regions in Belgium.

\parab{}($3$) The $20^{th}$ percentile of power generation for the $4$ wind farms above for the entire year of $2015$ at $1$~hour granularity as recorded in EMHIRES. These long-term thresholds (Iceland: $29$~MW, Norway: $16.5$~MW, Switzerland: $7$~MW, UK: $13.25$~MW) are different across sites due to local conditions and decide the provisioned compute cluster size. We deploy compute (in simulation) in multiples of NVIDIA H$100$ SuperPODS ($1$,$016$ GPUs, $1.3$~MW peak power draw).

\begin{figure*}[tbh]
    \centering
    \begin{minipage}{\textwidth}
        \centering
        \includegraphics[width=0.24\linewidth]{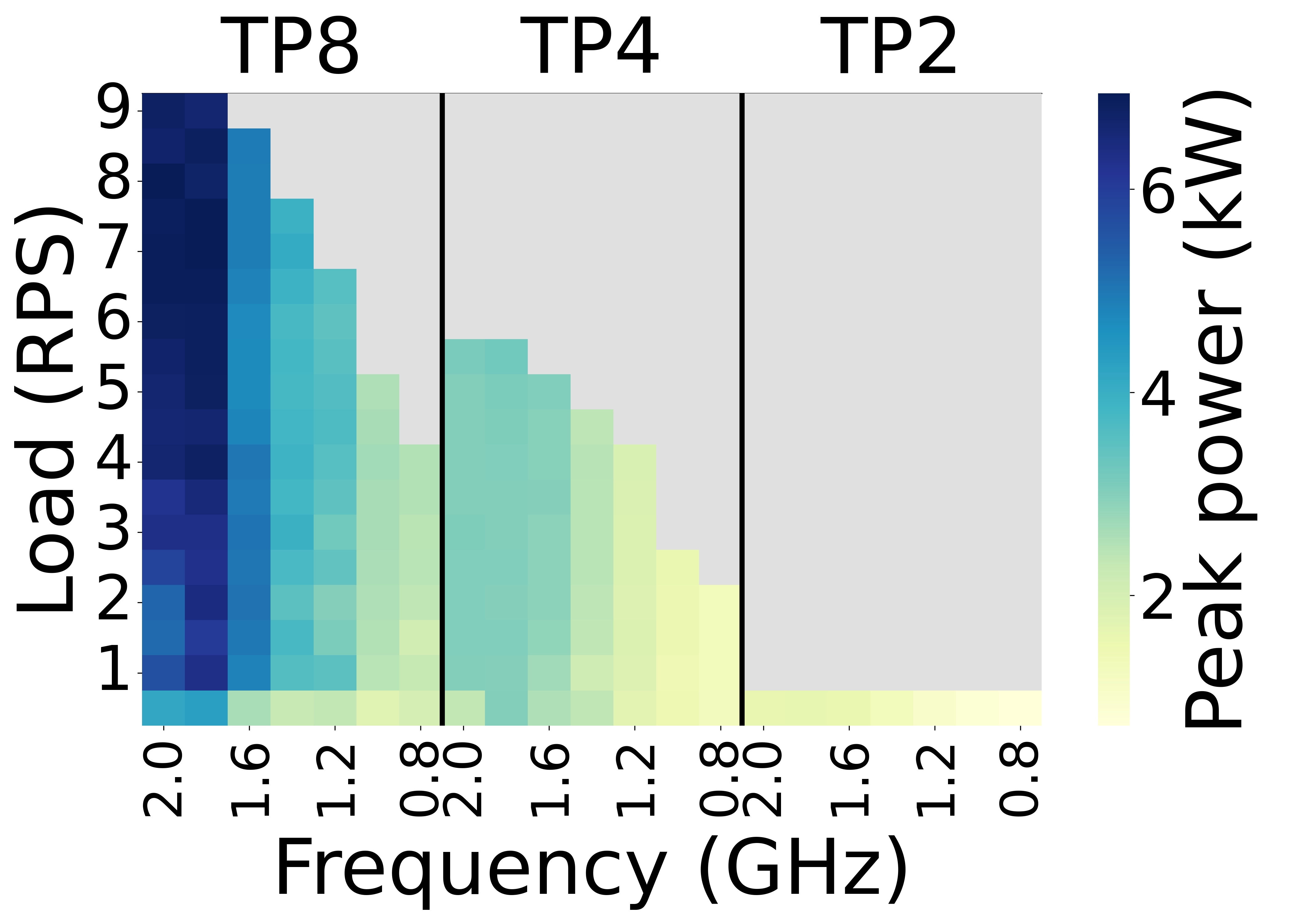}
        \hfill
        \includegraphics[width=0.24\linewidth]{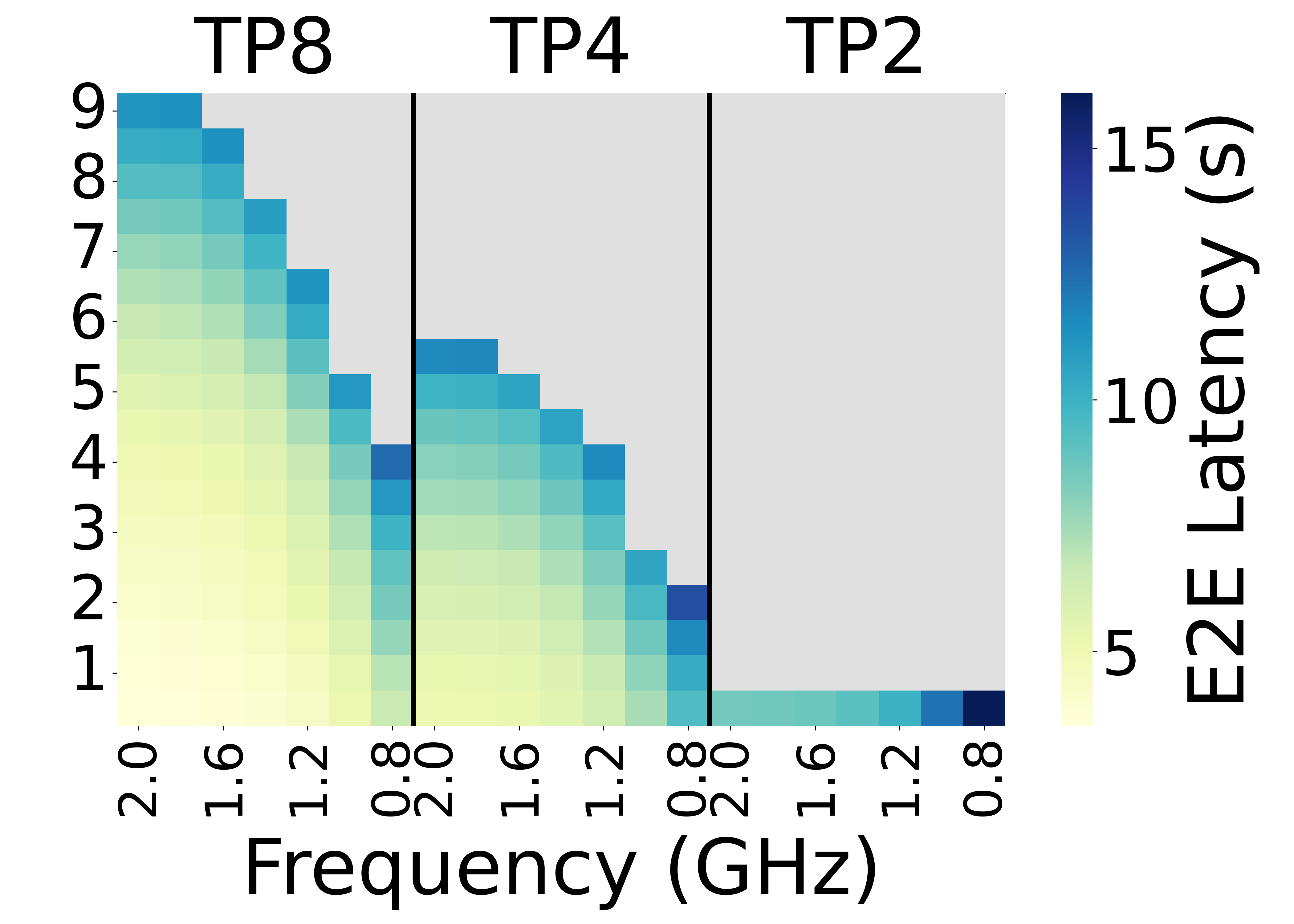}
        \hfill
        \includegraphics[width=0.24\linewidth]{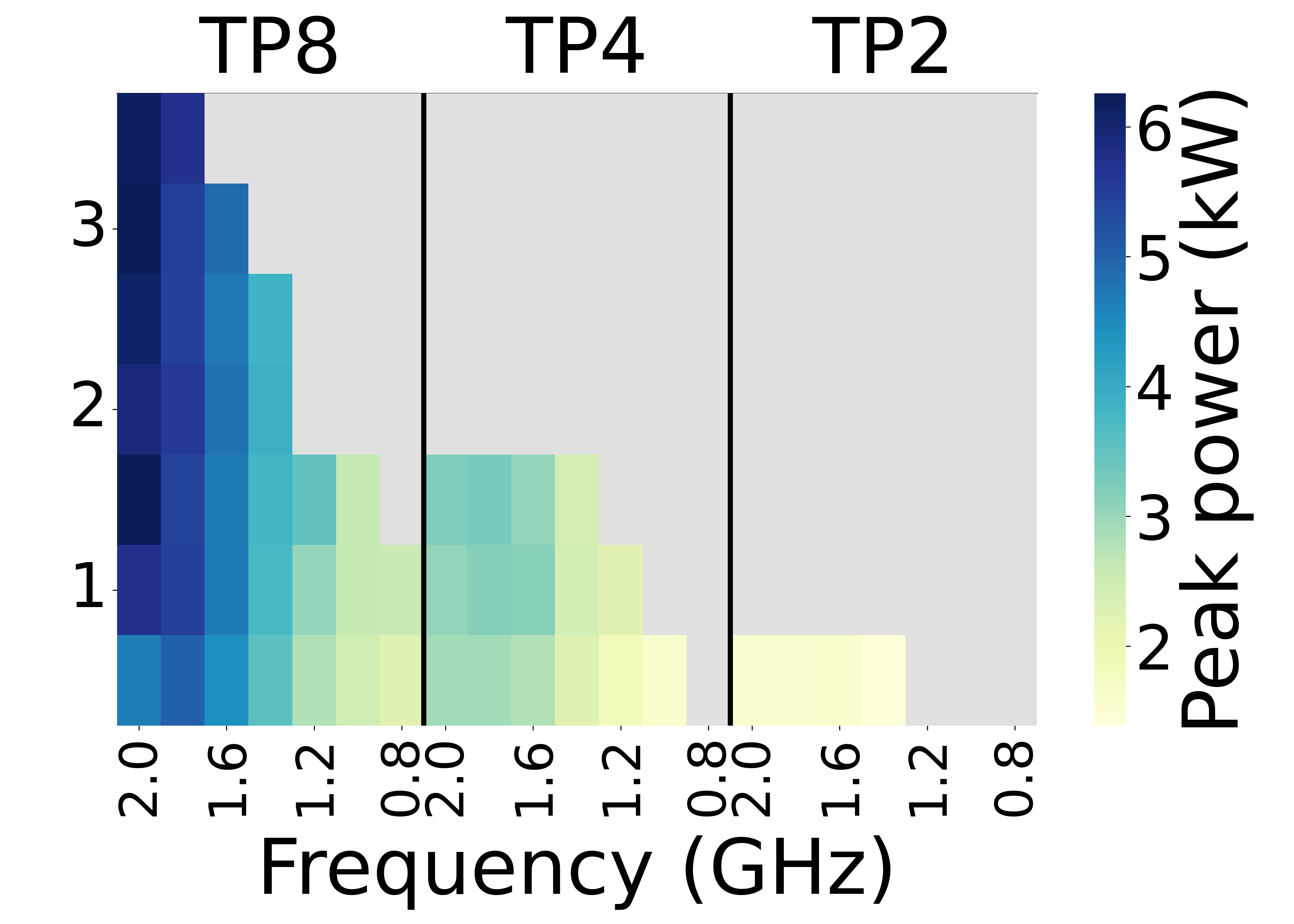}
        \hfill
        \includegraphics[width=0.24\linewidth]{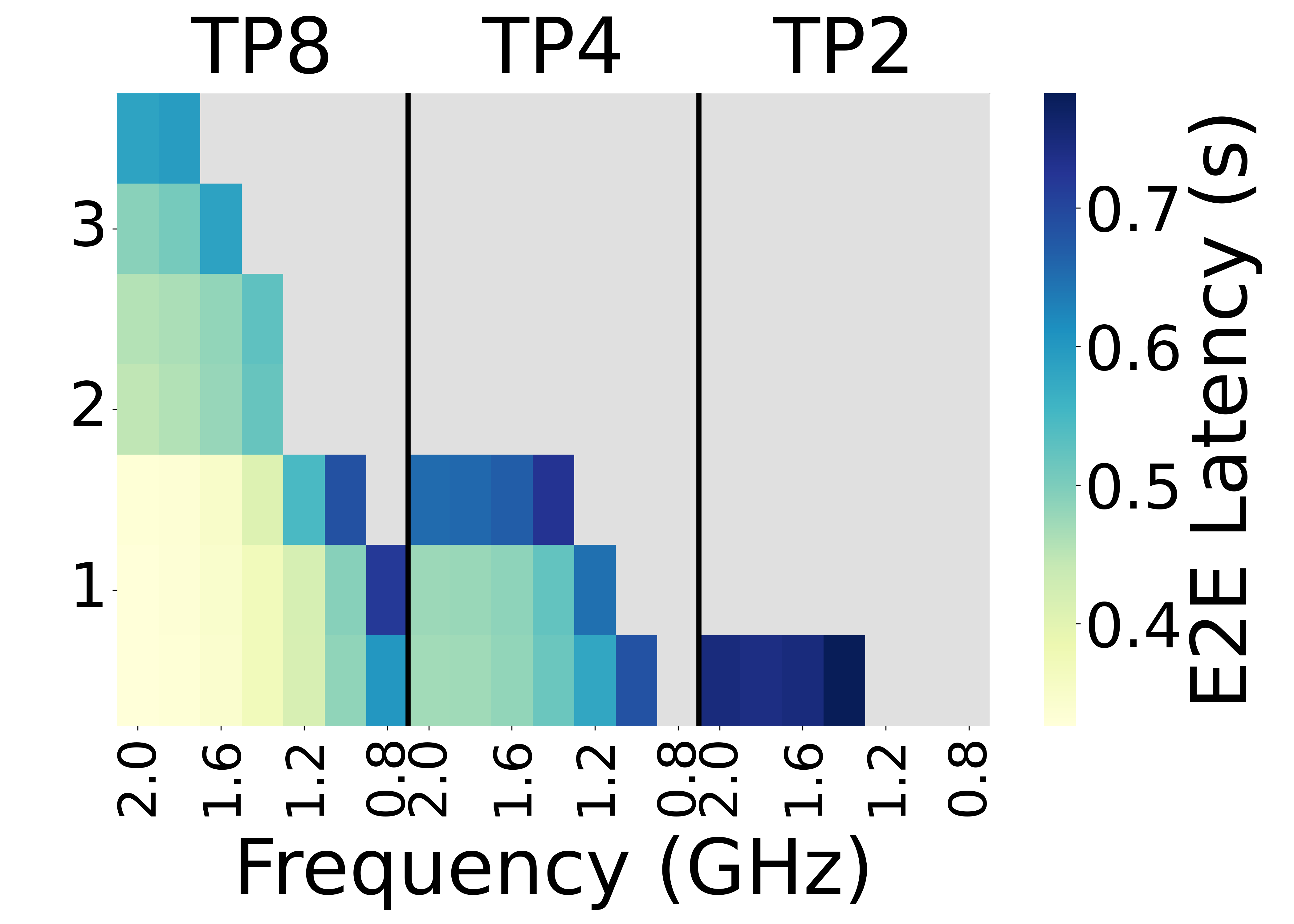}
    \end{minipage}
    \caption{Peak power and E$2$E latency for various load levels (requests per second or RPS) of MM requests across different TP and frequency combinations for \textit{conversation} (left) and \textit{coding} (right) workloads.}
    \label{fig:profiling_mm}
\end{figure*}

\subsection{Profiling The Workload}
\label{subsec:results:profiling} 

\noindent For each production trace, we categorize the requests into $9$ buckets \{SS, SM, SL, MS, MM, ML, LS, LM, LL\}---$3${}$\times${}$3$ combinations of \textbf{S}hort ($33^{rd}$ percentile over the entire week), \textbf{M}edium ($66^{th}$~ptile), and \textbf{L}ong ($100^{th}$~ptile) inputs and outputs. We profile the peak power, average power, TTFT (time to first token), TBT (time between tokens), and E$2$E (end-to-end) latency of each class of requests on H$100$ DGX and Llama $3.1$ $70$B model
for different combinations of load (requests/s), tensor parallelism (TP$2$, TP$4$, and TP$8$), and GPU frequencies ($0.8$, $1$, $1.2$, $1.4$, $1.6$, $1.8$, and $2$~GHz). For each combination of class $c$, frequency $f$, TP $t$, and load $l$, the GPU instance (multiple GPUs; e.g.\ TP$8$ needs $8$) caters to a Poisson arrival of requests with load $l$ for $5$~min, and we extract the power consumption statistics with NVIDIA's DCGM interfarce~\cite{nvidia_dcgm} and the latency statistics from the vLLM~\cite{vllm} serving engine. The TTFT and TBT latency SLOs are set to $5\times$ the TTFT/TBT values when a single request from a particular class is run in isolation at TP$8$ and $2$~GHz frequency.

Fig.~\ref{fig:profiling_mm} shows the peak power and E$2$E latency of MM \textit{conversation} (left) and \textit{coding} (right) requests under varying loads for different combinations of TP and frequencies. The grey cells (top right corners) denote TTFT/TBT violations---note that TP$2$ cannot sustain any load higher than $0.5$ requests/s for this class of requests. For a specific TP and frequency, as the load increases, both peak power consumption and latency (E$2$E as well as TTFT) inflate. Keeping everything else constant, a higher TP degree (or a larger operating frequency) leads to better latencies at the cost of higher power consumption. We have verified that the trends are similar across different classes of the $2$ traces, while the exact thresholds are different. \textit{Coding} trace has larger input (context) in general and hence cannot sustain high load levels (SLO violations) as compared to \textit{conversation}.

The profiling exercise generates lookup tables (functions) like $e2e(c,f,t,l)$ and $power(c,f,t,l)$ for all request classes for both \textit{conversation} and \textit{coding} traces that are leveraged by \rname. Note that \rname considers only those \{$c$,$f$,$t$,$s$,$l$\} combinations that \textit{do not lead to TTFT/TBT SLO violations}. Also, we used a fixed multiplier, $1.82$, for all the power numbers from $power(c,f,t,l)$ in all the following experiments to accommodate other components (CPU, fans, networks, memory) in the mix. Each H$100$ DGX box ($8$ GPUs) consumes a peak power of $10.2$~KW, which is $1.82\times$ higher than the aggregate peak power of the $8$ GPUs ($8 \times 700$~W).

\subsection{Planner-L Component Testing}
\label{subsec:planner-L}

\noindent Now we see \texttt{Planner-L} in action in isolation (without \rname component \texttt{Planner-S} \& \texttt{Request Sche\-duler's packing heuristic}). This exercise demonstrates how having a global view (akin to SDN's control plane~\cite{sdn}) across multiple sites and the power-variability awareness helps \rname unleash performance beyond the current baselines that use site-local views and naive round-robin routing. It also helps quantify the incremental benefits of the other \rname components in \S\ref{subsec:eval:all_components}. For the following experiments, we assume that the \texttt{Planner-L} decisions are blindly followed by the \texttt{Request Scheduler} and the \texttt{Configurator}.

\parab{Setup.} We run a week-long simulation for both production traces being served on Llama $3.1$ $70$B parameter model across the $4$ wind sites in Europe. 
We compare $2$ variants ($a$ \& $b$ below) of \texttt{Planner-L} schedulers against $2$ baselines as follows: ($a$) Planner-L-min-power: objective is to minimize total power consumed. ($b$) Planner-L-min-latency: objective is to minimize aggregate E$2$E latency per $15$~min. ($c$) WRR (weighted round-robin) + DynamoLLM: DynamoLLM scheduler~\cite{dynamollm} tries to minimize power draw over time (energy). A WRR router forwards fractions of incoming requests assigned in proportion to the compute per site. ($d$) Greedy minimum latency baseline: a heuristic that attempts to achieve low latency by assigning TP$8$, highest operating frequency, and lowest load level to GPU instances. However, we observed that it fails to assign $\sim${}$33\%$ requests over time due to GPU capacity constraints in place. To address this limitation, we increase the assigned load per GPU instance, as needed, while still assigning TP$8$ and highest frequency. We observed that the latency vs. load curve results in a ``knee point''---the latency increases before such a point is small. 
We calculate the knee point for each class of requests and limit the load on each GPU instance to these knee points.

\begin{figure*}[tbh]
    \centering
    \begin{minipage}{\textwidth}
        \centering
        \includegraphics[width=0.33\linewidth]{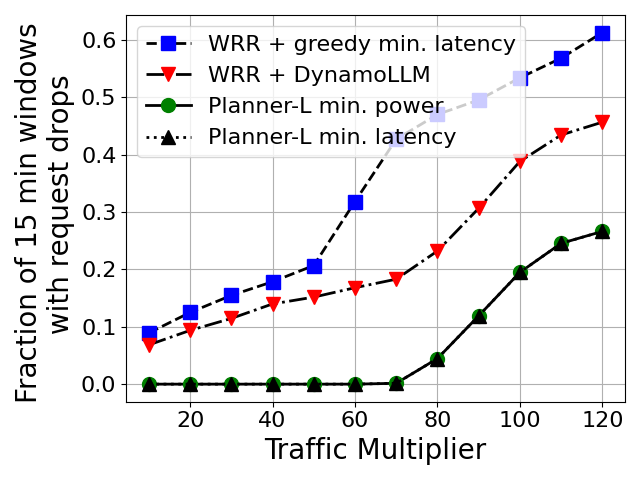}
        \hfill
        \includegraphics[width=0.33\linewidth]{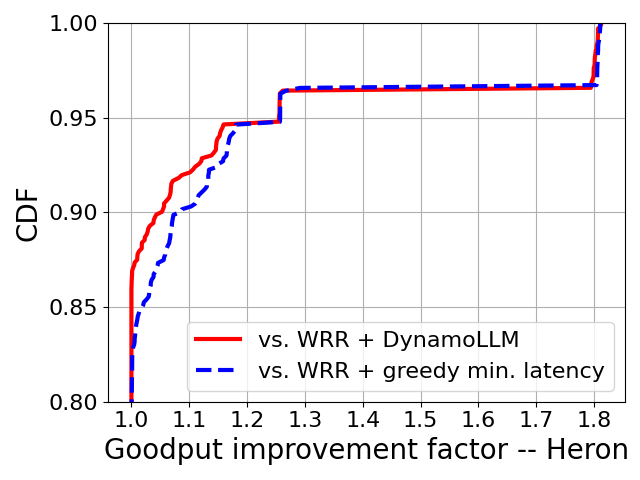}
        \hfill
        \includegraphics[width=0.33\linewidth]{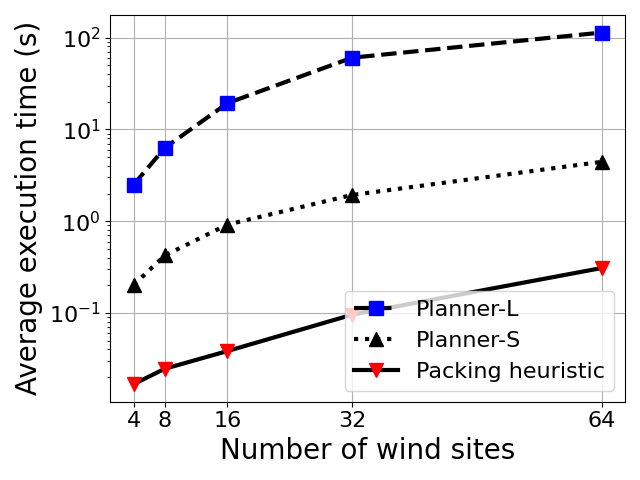}
    \end{minipage}
    \caption{(left) Across different volumes of \textit{coding} workload, the baselines lead to more frequent request drops. (middle) For a given workload, the goodput factor is significantly higher with \rname (planner-L) at higher percentiles (residual uncertainty). 
    (right) Execution times of \rname components.}
    \label{fig:planner-l}
\end{figure*}

\parab{\texttt{Planner-L} offers high goodput.} We first run the $4$ schedulers at $15$~min granularity and compare, in Fig.~\ref{fig:planner-l} (left), how frequently the schedulers hit corner cases with not enough power for the entire incoming workload. As neither DynamoLLM nor the minimum latency baseline is power-variability aware with WRR router not able to leverage power complementarity across sites, they onboard more workload per site than could be served leading to request drops. Fig.~\ref{fig:planner-l} (left) shows how often (number of $15$~min slots with at least one request drop) schedulers encounter request drops (due to local power unavailability) in a week for different volumes of the coding workload ($x$-times the arrival rate in the production traces). Both Planner-L variants (min. power and min. latency) could cater to significantly higher workload over time without any request drops. Essentially, planner-L min. latency converges to planner-L min. power in extreme resource-constrained cases -- if \texttt{Planner-L} min. power already drops requests due to resource constraints, the best plan that \texttt{Planner-L} min. latency could offer is same as the min. power one; hence, the overlap in the figure. Also, note that our choice of not using batteries, generators, or full-fledged DCs in the mix is stress-testing planner-L. In a real deployment, Planner-L could essentially guarantee no request drops, given its power-variability awareness. 

\begin{figure}[tbh]
    \centering
    \includegraphics[width=0.85\columnwidth]{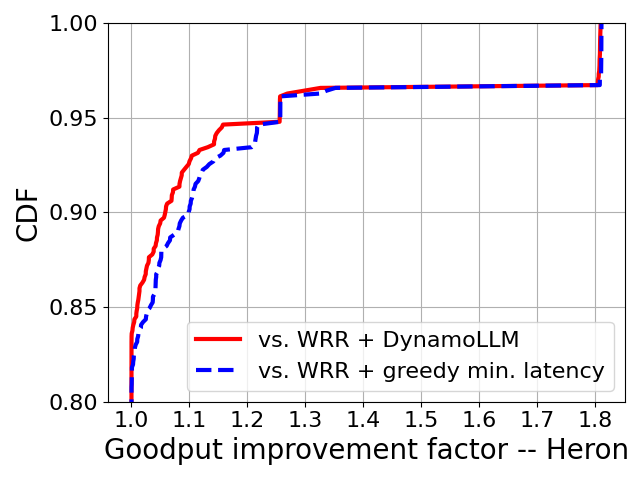}
    \caption{For \textit{conversation} ($50\times$) workload, the goodput factor is significantly higher with \rname (planner-L) at higher percentiles (residual uncertainty).}
    \label{fig:planner-l:goodput-conv-50x}
\end{figure}

Next, for a fixed multiplier of $60\times$ for the incoming \textit{coding} workload, we plot the goodput (requests being actually served) improvement factor for the Planner-L (both versions coincide) with respect to the $2$ baselines ($c$ and $d$). This improvement factor is the ratio of the goodput with Planner-L and the goodput with the corresponding baseline. Fig.~\ref{fig:planner-l} (middle) shows that planner-L (hence, \rname) could offer up to $1.8\times$ improvement in goodput compared to the baselines. Note that none of the $4$ choices have any difficulty serving the workload $80${}$+\%$ ($y=0.8$) of the time, thanks to our compute right-sizing strategy in \S\ref{sec:provisioning}. \textit{Even the baselines performing well up to a high-percentile in Fig.~\ref{fig:planner-l} (middle) is in fact one of our key contributions (deployment right-sizing discussed in \S\ref{sec:provisioning})}. The ratio stays at $1$ for a while even beyond the $80^{th}$ percentile mark in cases where the workload arrival rate at a site goes down as the power goes down. Also the mass at the top right ($80\%$ improvement in goodput) corresponds to the scenario when the Iceland site goes down completely at times---\rname (planner-L), in anticipation, intelligently assigns the workload to the other sites. 
Similarly, Fig~\ref{fig:planner-l:goodput-conv-50x} shows the goodput improvement with \rname (\texttt{Planner-L}) for the \textit{conversation} trace to be $50${}$\times$ as compared to the baselines with WRR routing under the same setup as in \S\ref{subsec:planner-L}. We see a $1.8${}$\times$ improvement for both variants of \texttt{Planner-L} none of which need to drop requests, thanks to \rname's capability to leverage spatial power generation complementarity.

\begin{figure}[t]
    \centering
    \begin{minipage}{0.5\columnwidth} 
        \centering
        \includegraphics[width=\textwidth]{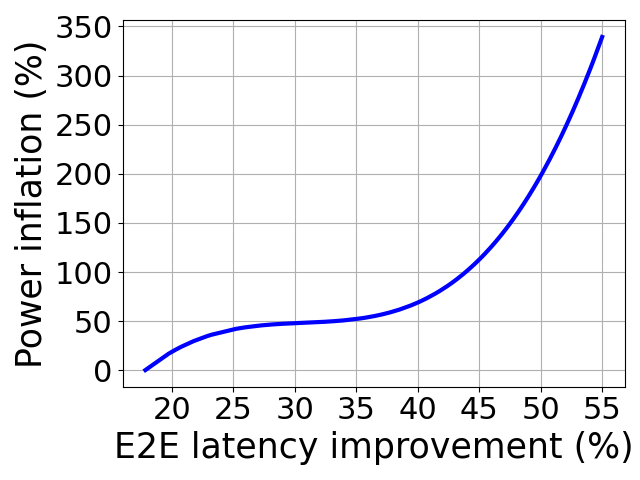}
    \end{minipage}%
    \hfill
    \begin{minipage}{0.45\columnwidth} 
        \caption{Planner-L min. latency offers lower latency at the cost of higher power than Planner-L min. power.}
        \label{fig:planner-l:tradeoff}
    \end{minipage}
\end{figure}

\parab{\texttt{Planner-L} variants offer a trade-off.} Fig.~\ref{fig:planner-l:tradeoff} demonstrates the trade-off offered by the $2$ variants of  Planner-L for \textit{coding} trace ($60${}$\times$) over the entire week. We fit a cubic curve to represent the \%-age improvement in E$2$E latency with Planner-L min. latency and the corresponding \%-age power inflation with respect to the min. power version (mean values per $15$~min slot over all requests). A $25\%$ (likewise, $50\%$) E$2$E latency improvement comes at the cost of $42\%$ (likewise, $198\%$) power inflation. For \textit{conversation} ($50${}$\times$) trace, a $25\%$ latency improvement corresponds to $49\%$ power inflation. 


\parab{\texttt{Planner-L} configuration stickiness.} The planner-L ILP uses a reconfiguration threshold $R_L$ ($X\%$) that limits the number of TP changes that are allowed per $15$~min slot (changes in generation are gradual in wind farms and span multiple such time-slots). We have run $1$-week experiments with different values of $R_L$ for both workloads, and have not seen any significant inflation in either the $95^{th}$-ptiles (across time-slots) of the mean (within a $15$~min slot) E$2$E latency or  the mean power consumed for $R_L$ values as low as $3\%$. Nevertheless, the latency inflates below $R_L=3\%$.

\parab{\texttt{Planner-L} scalability.} Fig.~\ref{fig:planner-l} (right) plots the execution time of \texttt{Planner-L} (alongside other \rname components that we discuss later) on a standard CPU for different number of wind-sites. Note that it increases linearly ($y$-axis is log-scale) in the range we tested for, always consuming $\leq1\%$ CPU and $\leq1$~GB RAM. Even for $64$ wind farm sites (cumulative power $2+$~GW), it executes in $6$~min which is way below the $15$~min budget. We used COIN-OR~\cite{coin-or} solver for our experiments -- the ILP could speed up, if needed, by running on commercial solvers like CPLEX~\cite{cplex} with multi-cores.

For all of the above results, the general observations and trends are similar across both \textit{coding} and \textit{conversation}. Hence, we focused on one of them reporting the other one as needed.

\begin{figure}[t]
    \centering
    \includegraphics[width=0.49\columnwidth]{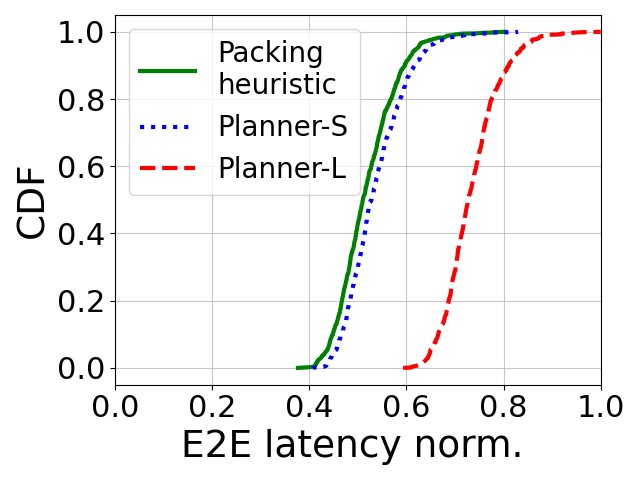}
    \includegraphics[width=0.49\columnwidth]{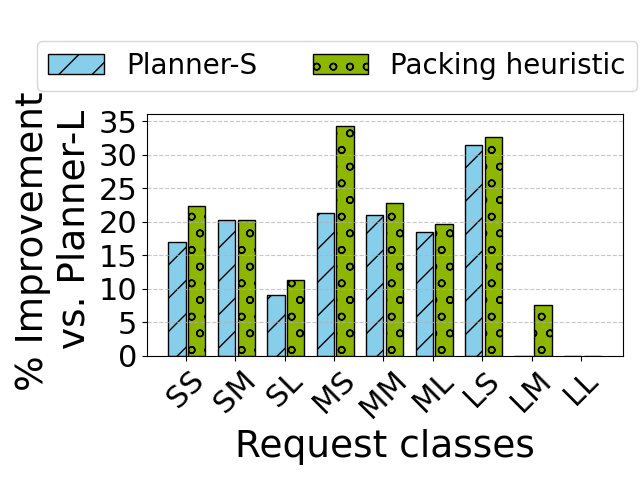}
    \caption{(left) Planner-S and the greedy packing heuristic incrementally improve E$2$E latency over the planner-L (min. latency) assignments. (right) Request class-wise mean E$2$E latency improvements over Planner-L.}
    \label{fig:heron_full:incremental}
\end{figure}

\subsection{Unleashing More Performance}
\label{subsec:eval:all_components}
\noindent \texttt{Planner-S} runs every $5$ seconds in this setup with $4$ wind farm sites reconfiguring operating frequencies and load as needed, and the \texttt{packing heuristic} (with \texttt{Request Sche\-duler}) runs online greedily packing smaller requests to GPU instances configured for larger ones (e.g. LS requests could be assigned to LM instance) leading to E$2$E latency improvements. In Fig.~\ref{fig:heron_full:incremental} we show the incremental latency benefits of running these components in tandem with \texttt{Planner-L} (min. latency) for a $15$~min slot. The CDF plots (Fig.~\ref{fig:heron_full:incremental}, left) show the E$2$E latencies (mean per second) when requests are routed based on \texttt{Planner-L}, \texttt{Planner-L+S}, and \texttt{Planner-L+S+packing heuristic} assignments. At the median, the $E2E$ latency improvements in aggregate are $27\%$ with \texttt{Planner-S} and an additional $3\%$ with the \texttt{packing heuristic} compared to only \texttt{Planner-L}. For the \textit{conversation} trace, \texttt{Planner-S} offers $12\%$ improvement (at the median) in latency over \texttt{Planner-L} min. latency and the \texttt{packing heuristic} offers a further $2\%$ improvement. Fig.~\ref{fig:heron_full:incremental}  (right) shows the request class-wise mean reductions in latency over \texttt{Planner-L}---the incremental improvements vary between $7.5\%$ and $34.4\%$ with LL requests seeing no improvements as the \texttt{packing heuristic} does not have a better class to assign these requests to. Note that both \texttt{Planner-L} and \texttt{Planner-S} are run with the same optimization goal -- in this case, minimizing the E$2$E latency.

\parab{\texttt{Planner-S} power elasticity.} We stress-tested \texttt{Planner-L} by reducing the available power by $20\%$ and still routing as per \texttt{Planner-L} assignments. We picked a $5$~s window at random and observed that, only for one $1$~s window, $\sim${}$7\%$ requests had to be dropped to be within the reduced power budget (than anticipated). \texttt{Planner-S} though could accommodate all requests for this $1$~s window. 

\parab{Scalability.} As seen in Fig.~\ref{fig:planner-l} (right), \texttt{Planner-S} is $\sim${}$30${}$\times$ faster than \texttt{Planner-L}. For smaller setups with $4$ wind sites, \texttt{Planner-S} could run multiple times in a second, and for larger $64$ site deployments it could run every $5$~seconds ($180$ times in $15$~min). Hence, \texttt{Planner-S} could be multiple orders of magnitude more fine-grained than \texttt{Planner-L}, also because it does not touch the TP configurations. The \texttt{packing heuristic} is, again, significantly quicker (sub-second even for larger deployments) than the other components and could run with the online \texttt{Request Scheduler}.

\section{Related Work}
\label{sec:related_work}

\parab{DC running on renewables.} Previous works~\cite{parasol, zccloud, greenpar, singh2013yank, li2012iswitch} explored renewable-powered DCs leveraging solar power, battery, grid draw, and scheduling optimization. 
Key differences with AI Greenferencing are: ($1$) \textit{Driver:} We aim to unlock more AI compute motivated by the growing power demand. ($2$) \textit{Power source:} We focus on wind energy as solar generation needs to rely on batteries or generators heavily. ($3$) \textit{Right-sizing:} We recommend smaller compute in larger wind farms, a strategy that significantly reduces power uncertainty. ($4$) \textit{Cross-site:} Our work leverages power complementarity across sites. ($5$) \textit{Workload:} AI inferencing is more amenable to power fluctuations than traditional workloads.

\parab{Moving compute to power.} Virtual Battery \& SkyBox~\cite{virtual_battery, skybox} use renewable power complementarity across sites to run compute. Differences with AI Greenferencing: ($1$) \textit{Provisioning:} We focus on compute right-sizing and intelligent routing software alongside power complementarity. ($2$) \textit{Scalability:} While SkyBox does not leverage sites with large variations in generation, we could still include them. ($3$) \textit{Workload:} SkyBox focuses on heavier VM placement and migration. We rather focus on intelligent routing of AI inferencing requests.


\parab{Power-aware AI Inferencing.} DynamoLLM~\cite{dynamollm} minimizes energy/power consumption in AI inferencing by tuning TP degree and GPU frequencies. While we build \rname on top of this work, the key differences are: ($1$) \textit{Cross-site:} Our work focuses on cross-site routing and not site-specific scheduling. ($2$) \textit{Assumptions:} DynamoLLM assumes no power or compute limitations, while we are constrained on both at wind sites. ($3$) \textit{Approach:} We are power-variability aware. 

\section{Discussions}
\label{sec:discussions}


\parab{Batteries/generators.} They help alleviate some of the wind power uncertainty at the cost of round-trip loss, hybrid energy mix, and higher CAPEX. Here, we stress-tested \rname without using any of these additional modalities and rather leverage right provisioning, power complementarity, and application/infrastructure knobs to tackle power uncertainty.

\parab{Beyond wind.} We leave exploring other sustainable energy sources for AI workloads as future work. Solar farms face $\sim${}$50\%$ unavailability, requiring costly battery/generator backup, increasing CAPEX and OPEX. While nuclear energy is on the radar of big AI players~\cite{google_nuclear, ms_three_mile} for its stable base load, it has its own challenges -- ($1$) high CAPEX, ($2$) nuclear waste management, ($3$) public perception and regulations, ($4$) restricted zones, and ($5$) long setup times.

\parab{GPU phase-out.} With GPU leaders like NVIDIA announcing a new generation almost every year, 
hyperscalars need 
concrete plans to phase-out the quickly aging GPUs. AI Greenferencing could help ship these GPUs to wind farms and run a fraction of the relevant AI workload at low CAPEX.

\parab{Autonomous maintenance.} Alongside training farm-local labor on basic cluster maintenance, one could also engage cheap robots in these MDCs for basic repairing tasks~\cite{hong2024self}.

\parab{Elephants versus mice.} In AI Greenferencing, a large wind farm paired with smaller deployments might cause imbalances, requiring reroutes beyond the smaller sites’ capacity. Fig.~\ref{fig:connectivity:global_map} shows DCs have diverse wind farms within low network latency, enabling sizeable deployments to handle power uncertainty effectively.




\parab{Security of the models.} An AI inferencing provider can either deploy their own compute or leverage $3^{rd}$ party compute (startups~\cite{soluna, windcores, westfalenwind} co-locating compute in wind farms). With the latter approach, the AI provider needs to resort to privacy-preserving model-splitting techniques~\cite{fu2024privacy, guo2021lightfed}. 


\section{Conclusion} 

\noindent AI Greenferencing brings AI compute to wind energy sources, bypassing power delivery challenges and meeting rising AI demand sustainably. We present a feasibility study and a right-sizing strategy to mitigate wind power uncertainties. Our cross-site router efficiently manages residual uncertainties, delivering up to $1.8${}$\times$ the goodput of state-of-the-art solutions. \textit{This work does not raise any ethical issues.}


\bibliographystyle{ACM-Reference-Format}
\bibliography{paper}


\end{document}